\documentclass[fleqn,usenatbib]{mnras}

\usepackage{newtxtext,newtxmath}
\usepackage{xspace}
\usepackage[T1]{fontenc}
\usepackage{ae,aecompl}

\usepackage{graphicx}	
\usepackage{amsmath}	
\usepackage{comment}
\usepackage{multirow}
\usepackage{paralist}

\newcommand{\code}[1]{\textsc{#1}\xspace}

\newcommand{\wit}{\mbox{VVV-WIT-08}}
\newcommand{\Ks}{$K_s$}

\newcommand{\VVVx}{VVVX}
\newcommand{\gaia}{\textit{Gaia}}

\title[VVV-WIT-08]{VVV-WIT-08: the giant star that blinked}

\author[L. C. Smith et al.]{
Leigh C. Smith$^{1}$\thanks{E-mail: lsmith@ast.cam.ac.uk},
Sergey E. Koposov$^{2,1,3}$,
Philip W. Lucas$^{4}$,
Jason L. Sanders$^{5,1}$,
\newauthor
Dante Minniti$^{6,7}$,
Andrzej Udalski$^{8}$,
N. Wyn Evans$^{1}$,
David Aguado$^{1}$,
\newauthor
Valentin D. Ivanov$^{9}$,
Roberto K. Saito$^{10}$,
Luciano Fraga$^{11}$,
Pawel Pietrukowicz$^{8}$,
\newauthor
Zephyr Penoyre$^{1}$,
Carlos Gonz\'alez-Fern\'andez$^{1}$
\\
$^{1}$Institute of Astronomy, University of Cambridge, Madingley Rd, Cambridge, CB3 0HA, UK\\
$^{2}$Institute for Astronomy, University of Edinburgh, Royal Observatory, Blackford Hill, Edinburgh EH9 3HJ, UK\\
$^{3}$McWilliams Center for Cosmology, Carnegie Mellon University, 5000 Forbes Ave, 15213\\
$^{4}$Centre for Astrophysics, University of Hertfordshire, College Lane, Hatfield AL10 9AB, UK\\
$^{5}$Department of Physics and Astronomy, University College London, London WC1E 6BT, UK\\
$^{6}$Departamento de Ciencias F\'isicas, Universidad Andr{\'e}s Bello, Fern{\'a}ndez Concha 700, Las Condes, Santiago, Chile\\
$^{7}$Vatican Observatory, V00120 Vatican City State, Italy\\
$^{8}$Astronomical Observatory, University of Warsaw Al. Ujazdowskie 4,00-478 Warszawa, Poland\\
$^{9}$European Southern Observatory, Karl-Schwarzschild-Str. 2, 85748, Garching bei M\"unchen, Germany\\
$^{10}$Departamento de  F\'{i}sica, Universidade  Federal de  Santa Catarina, Trindade 88040-900, Florian\'opolis, SC, Brazil\\
$^{11}$Laborat\'orio Nacional de Astrof\'isica LNA/MCTI, 37504-364 Itajub\'a, MG, Brazil\\
}

\date{Accepted XXX. Received YYY; in original form ZZZ}

\pubyear{2020}

\begin{document}
\label{firstpage}
\pagerange{\pageref{firstpage}--\pageref{lastpage}}
\maketitle

\begin{abstract}
We report the serendipitous discovery of a late-type giant star that exhibited a smooth, eclipse-like drop in flux to a depth of $97$ per cent. Minimum flux occurred in April 2012 and the total event duration was a few hundred days. Light curves in $V$, $I$ and \Ks{} from the Optical Gravitational Lensing Experiment and VISTA Variables in the Via Lactea surveys show a remarkably achromatic event. During 17 years of observational coverage of this source only one such event was detected. The physical properties of the giant star itself appear somewhat unusual, which may ultimately provide a clue towards the nature of the system. By modelling the event as an occultation by an object that is elliptical in projection with uniform transparency, we place limits on its physical size and velocity. We find that the occultation is unlikely to be due to a chance alignment with a foreground object. We consider a number of possible candidates for the occulter, which must be optically thick and possess a radius or thickness in excess of $0.25$~au. None are completely satisfactory matches to all the data. The duration, depth and relative achromaticity of the dip mark this out as an exceptionally unusual event, whose secret has still not been fully revealed. We find two further candidates in the VVV survey and we suggest that these systems, and two previously known examples, may point to a broad class of long period eclipsing binaries wherein a giant star is occulted by a circumsecondary disc.
\end{abstract}




\section{Introduction}

$\epsilon{}$ Aurigae is a binary star on a 27 year orbit. The primary is a F0 type star with supergiant gravity thought to be the result of a late thermal flash; the B-type secondary hosts a large and opaque disc. The eclipse of the primary by the secondary causes a long ($\sim{}2$~year), deep ($\Delta{_{V {\rm mag}}}\approx{}1$), flat-bottomed dip in the light curve. \citet{ea10, ea15} presented the results of their interferometric observations of the 2009-2011 eclipse, including superb reconstructed images of the occulting disc in silhouette.

\citet{Denisenko13} identified the ``optical anti-transient'' TYC 2505-672-1, and further work by \citet{Lipunov16} and \citet{Rodriguez16} presented it as a more extreme eclipsing binary system than $\epsilon{}$ Aurigae, which had previously held the record for longest orbital period. TYC 2505-672-1 is a binary system on a 69 year orbit, comprising an M-type giant primary with a proposed B-type sub-giant companion hosting a large, opaque circumstellar disc. The flat-bottomed eclipse lasts $\sim{}3.45$ years and reaches a depth of $\Delta{_{V {\rm mag}}}\approx{}4.5$.

A third example of a long-period eclipsing binary analogous to $\epsilon{}$ Aurigae and TYC~2505-672-1 is ASASSN-21co \citep{asassn-21co}, an M-type giant star which began an eclipse event in February 2021. A previous event that lasted $\sim{}80$~days was observed as part of the All Sky Automated Survey \citep{asas} in 2009, suggestive of a 12 year period. The current event, ongoing at the time of preparation of this article, is therefore likely to continue until the end of April 2021.

When it comes to the modelling of occultation events, V1400 Centauri (also known as Mamajek's Object) springs to mind. It is a K5 pre-main sequence star which exhibited a long, deep and complex dipping event in SuperWASP photometry in 2007. \citet{Mamajek12} found the light curve was consistent with an occultation by a large ($\approx{}0.4$ au) and complex disc and ring system hosted by a low mass companion orbiting the primary.

Another a remarkable occultation event was identified and modelled by \citet{Rappaport19}, who also provide a useful list of dippers of various types in the literature. It is a single $80$ per cent depth event of width 1 day in the light curve of the young, nearby M dwarf star EPIC~204376071. The dip is notably smooth and asymmetric, a trait which the authors ascribe to an occultation by an inclined and tilted disc likely hosted by a second body in orbit around the M dwarf.

The VISTA Variables in the Via-Lactea (VVV; \citealt{vvv10}) survey, and its extension the VVV eXtended (VVVX; \citealt{vvvx}) survey, sample $>10^9$ sources in the southern Galactic disc and bulge over hundreds of epochs across nearly a decade. It is not surprising that among them we find many unique variable objects exhibiting extreme behaviour that does not fit any known class of stellar variability. These special variable sources have been dubbed \lq\lq what-is-this", or \lq\lq WIT" objects, and follow-up investigations have revealed the nature of a few of them. The list of \lq\lq WIT" objects includes:
\begin{description}
\item VVV-WIT-01, probably an obscured nova or a protostellar collision \citep{wit01};
\item VVV-WIT-04, an extragalactic source thought to be an infrared violent variable quasar \citep{wit04};
\item VVV-WIT-06, a transient of unknown nature that may be a rare Galactic nova, or a stellar merger  \citep{wit06_minni,wit06_baner};
\item and VVV-WIT-07, an intriguing variable source thought to fit within the dipper paradigm  \citep{wit07}.
\end{description}

Here, we present another unusual transient object -- \wit{} -- exhibiting a single smooth, near-symmetrical dipping event over approximately $200$~days. The event is largely achromatic across optical and near-infrared wavelengths. It is at first glance reminiscent of a grazing exoplanet transit, but it has a depth of $\approx{}97$ per cent in the available wavebands.

Subsequent shallow searches in the VVV and VVVX data revealed two other promising candidate events exhibiting similar deep reductions in flux, albeit with poorer light curve sampling, that could not be readily explained through common causes of stellar variability. Since their data are not as complete as for \wit{} we postpone their analysis and a more comprehensive search for other similar events for a future work.

In Section \ref{sec:data} we summarise the data we have collected for \wit{}. We estimate the physical parameters of \wit{} in Section \ref{sec:physical_params}. Section \ref{event_modelling} analyses the event itself and describes our attempt to model it, where Section \ref{discussion} discusses how the models compare to the parameters of known objects. We draw conclusions in Section~\ref{conclusion}.

\section{Data}\label{sec:data}

\wit{} was initially identified during validation of a preliminary VVV and \VVVx{} light curve database and separately in a search for $\Delta{K_s}>4$~mag high amplitude variable stars\footnote{The final measured $K_s$ amplitude was a little under $4$ mag as the preliminary $\Delta{K_s}>4$~mag measurement included a poor quality observation at minimum flux which we've since discounted.} in that same database.

A further shallow, targeted search in the preliminary database yielded a second source at with a $\Delta{K_s}\approx{}3$~mag event, \mbox{VVV-WIT-10}. Subsequent to this, data processing and pipeline improvements led to a revised database and a further search for VVV variable stars and transients having $\Delta{K_s}>4$~mag (Lucas et al., in prep). This further search yielded a third source exhibiting a significant unexplained drop in flux, \mbox{VVV-WIT-11}.

The epoch J2016.0 coordinates of \mbox{VVV-WIT-10} are $\alpha=$~15h09m33.80s and $\delta=$~-59d38m42.56s, courtesy of Gaia eDR3\footnote{VVV-WIT-10 Gaia eDR3 source\_id~=~5876552066041521024} \citep{GeDR3}. It has $K_s=12.78$~mag and $J-K_s=2.01$~mag in the out-of-event photometry. Galactic coordinates are $l,b=319.58^{\circ},-1.32^{\circ}$, notably a disc field.

The epoch J2014.0 coordinates of \mbox{VVV-WIT-11} are $\alpha=$~17h33m41.26s and $\delta=$~-35d44m16.74s, courtesy of the VVV survey. \mbox{VVV-WIT-11} is present in Gaia eDR3\footnote{VVV-WIT-11 Gaia eDR3 source\_id~=~5974812366541771520}, but the VVV position is more precise due to the source having a magnitude of $20.44$ in the Gaia $G$ band. It has $K_s=12.15$~mag and $J-K_s=2.38$~mag in the out-of-event photometry. Galactic coordinates are $l,b=352.90^{\circ},-1.49^{\circ}$.

The discovery of \mbox{VVV-WIT-11} was fortuitous, as the source disappears completely for an entire observing season. As a result it would not ordinarily have been selected as part of the $\Delta{K_s}>4$~mag search (which did not include upper limits), were it not for a handful of mismatches with an adjacent slightly blended faint star during the dimming event. The absence of a detection during the dimming event suggests $\Delta{K_s}\gtrsim{}5$~mag.

The multi-band time series photometry for \mbox{VVV-WIT-08}, \mbox{VVV-WIT-10}, and \mbox{VVV-WIT-11} are included in the supplementary material. Also included is a figure showing the VVV \Ks{} and OGLE $I$ band light curve of \mbox{VVV-WIT-10}.

A summary of the photometric and other data we have collected for \wit{} is provided in Table \ref{data} and these data are described in more detail below.

{\renewcommand{\arraystretch}{1.1}
\begin{table}
\caption{Data for \wit{} obtained from a variety of sources. Photometric errors, particularly ground based ones, are likely to be subject to additional absolute calibration errors.}
\label{data}
\begin{center}
\resizebox{\columnwidth}{!}{
\begin{tabular}{|c|c|l|c|}
\hline
  \multicolumn{1}{|c|}{Parameter} &
  \multicolumn{1}{c|}{Value} &
  \multicolumn{1}{c|}{Unit} &
  \multicolumn{1}{c|}{Source} \\
\hline  

ID & 4044152029396602880 & & \multirow{9}{*}{Gaia eDR3 $^a$} \\
$\alpha_{2016.0}$ &  18h00m47.41s & &  \\
$\delta_{2016.0}$ & -30d36m09.28s & & \\
$\pi$ & $0.0091\pm0.0737$ &mas &  \\
$\mu_{\alpha\cos\delta}$ & $-3.670\pm0.081$ &mas~yr$^{-1}$ &  \\
$\mu_{\delta}$ & $-6.981\pm0.056$ &mas~yr$^{-1}$ &  \\
$G_\mathrm{BP}$ & $17.3964\pm0.0139$ & mag & \\
$G$ & $16.0340\pm0.0010$ & mag &  \\
$G_\mathrm{RP}$ & $14.7494\pm0.0237$ & mag &  \\

\hline

$r$ & $16.292\pm0.004$ & mag & \multirow{3}{*}{VPHAS$+$ $^b$} \\
$i$ & $14.974\pm0.003$ & mag &  \\
$H\alpha$ & $15.776\pm0.005$ & mag & \\

\hline

$g$ & $18.267\pm0.009$ & mag & \multirow{4}{*}{PS1 $^c$} \\
$i$ & $15.354^i$         & mag & \\
$z$ & $14.768\pm0.001$ & mag & \\
$y$ & $14.410\pm0.003$ & mag & \\

\hline

$g$ & $18.290\pm0.006$ & mag & \multirow{5}{*}{DECaPS $^d$} \\
$r$ & $16.200\pm0.006$ & mag & \\
$i$ & $15.145\pm0.006$ & mag & \\
$z$ & $14.506\pm0.007$ & mag & \\
$Y$ & $14.234\pm0.006$ & mag & \\

\hline

$Z$ & $14.100\pm0.013$ & mag & \multirow{5}{*}{VVV $^e$} \\
$Y$ & $13.448\pm0.013$ & mag & \\
$J$ & $12.778\pm0.003$ & mag & \\
$H$ & $11.842\pm0.006$ & mag & \\
$K_{s}$ & $11.500\pm0.002$ & mag & \\

\hline

$J$ & $12.767\pm0.035$ & mag & \multirow{3}{*}{2MASS $^f$} \\
$H$ & $11.736\pm0.031$ & mag & \\
$K_{s}$ & $11.421\pm0.027$ & mag & \\

\hline

$W1$ & $11.118\pm0.031$ & mag & \multirow{4}{*}{AllWISE $^g$} \\
$W2$ & $11.525\pm0.028$ & mag & \\
$W3$ & $>11.668$ & mag & \\
$W4$ & $>8.834$ & mag & \\

\hline

$3.6\,\mu$m & $11.292\pm0.048$ & mag & \multirow{4}{*}{GLIMPSE $^h$} \\
$4.5\,\mu$m & $11.255\pm0.064$ & mag & \\
$5.8\,\mu$m & $11.278\pm0.058$ & mag & \\
$8.0\,\mu$m & $11.193\pm0.045$ & mag & \\

\hline
\end{tabular}
}
\end{center}
$^a$ \citet{Gaia16, GeDR3}.\\
$^b$ \citet{vphas}.\\
$^c$ \citet{PS1}.\\
$^d$ \citet{decaps18}.\\
$^e$ \citet{vvv10}.\\
$^f$ \citet{2MASS}.\\
$^g$ \citet{wise10, allwise13}.\\
$^h$ \citet{glimpse09}.\\
$^i$ No error given due to this measurement being based on one observation.
\end{table}
}

\subsection{Light curves}

A detailed description of the VVV and VVVX light curve database and data processing is outside the scope of this paper. In summary, VVV and \VVVx{} images were processed with a modified version of the DoPHOT profile fit image reduction software \citep{dophot1, dophot2}. The resultant photometric measurements were then calibrated through a globally optimized zero point plus illumination map model, and a further high spatial frequency noise reduction routine is used to reduce scatter in the light curves. Both photometric calibration stages are described in detail in Smith et al. (in prep). VVV and \VVVx{} observations cover the 2010-2019 calendar years.

The Optical Gravitational Lensing Experiment (OGLE) survey covers the area of sky containing \wit{} as part of OGLE-III and -IV \citep{ogleiii, ogleiv} and provides data from 2001 to 2019. Due to heavy blending and huge brightness variations of \wit{}, the standard pipeline photometry was insufficient and a re-reduction and calibration was necessary. \wit{} was still faintly visible in images taken during the peak of the dimming. To produce a high quality light curve at minimum flux, we selected good-seeing in-event images and constructed new templates for a dedicated image subtraction in the $I$ and $V$ bands.

\begin{figure*}
  \begin{center}
    \includegraphics[width=\textwidth,keepaspectratio]{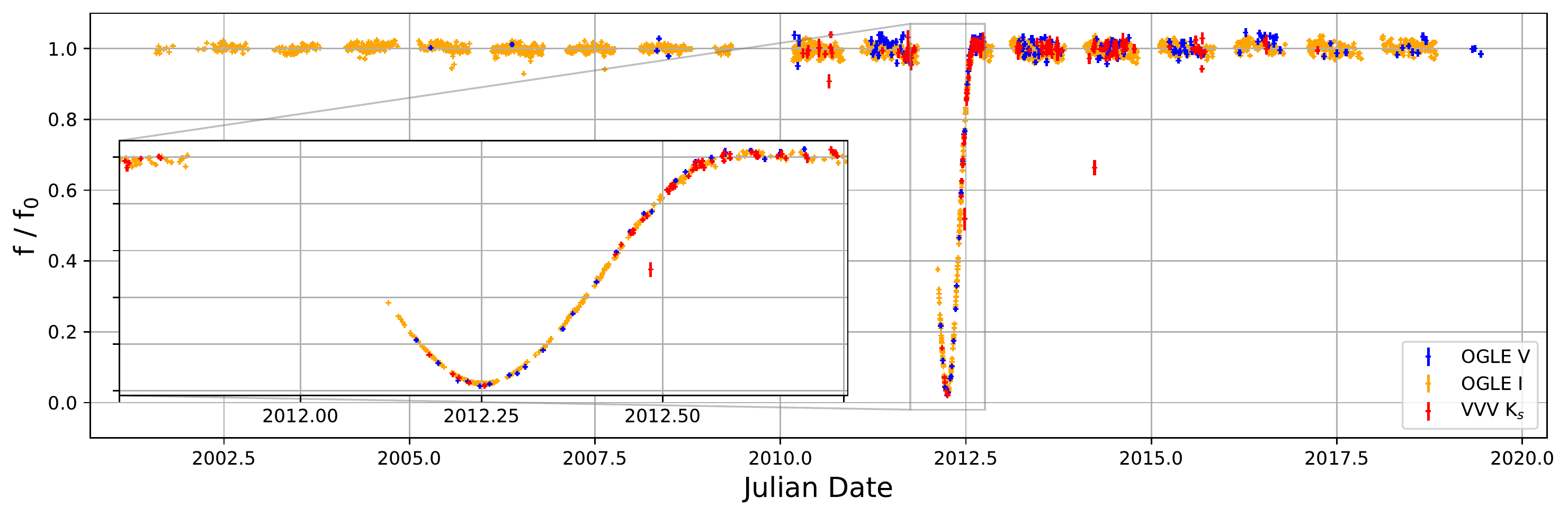}\\
    \caption{OGLE $V$ and $I$ bands and VVV and \VVVx{} \Ks{} band flux relative to the inverse variance weighted mean of the flux outside the event ($|t-t_0|>365$~days) for \wit{}. Inset is a closer view of $1$ year of data centered on the minimum $I$ band flux measurement.}
    \label{fig:VIKs_lc}
  \end{center}
\end{figure*}

\begin{figure}
  \begin{center}
    \includegraphics[width=\linewidth,keepaspectratio]{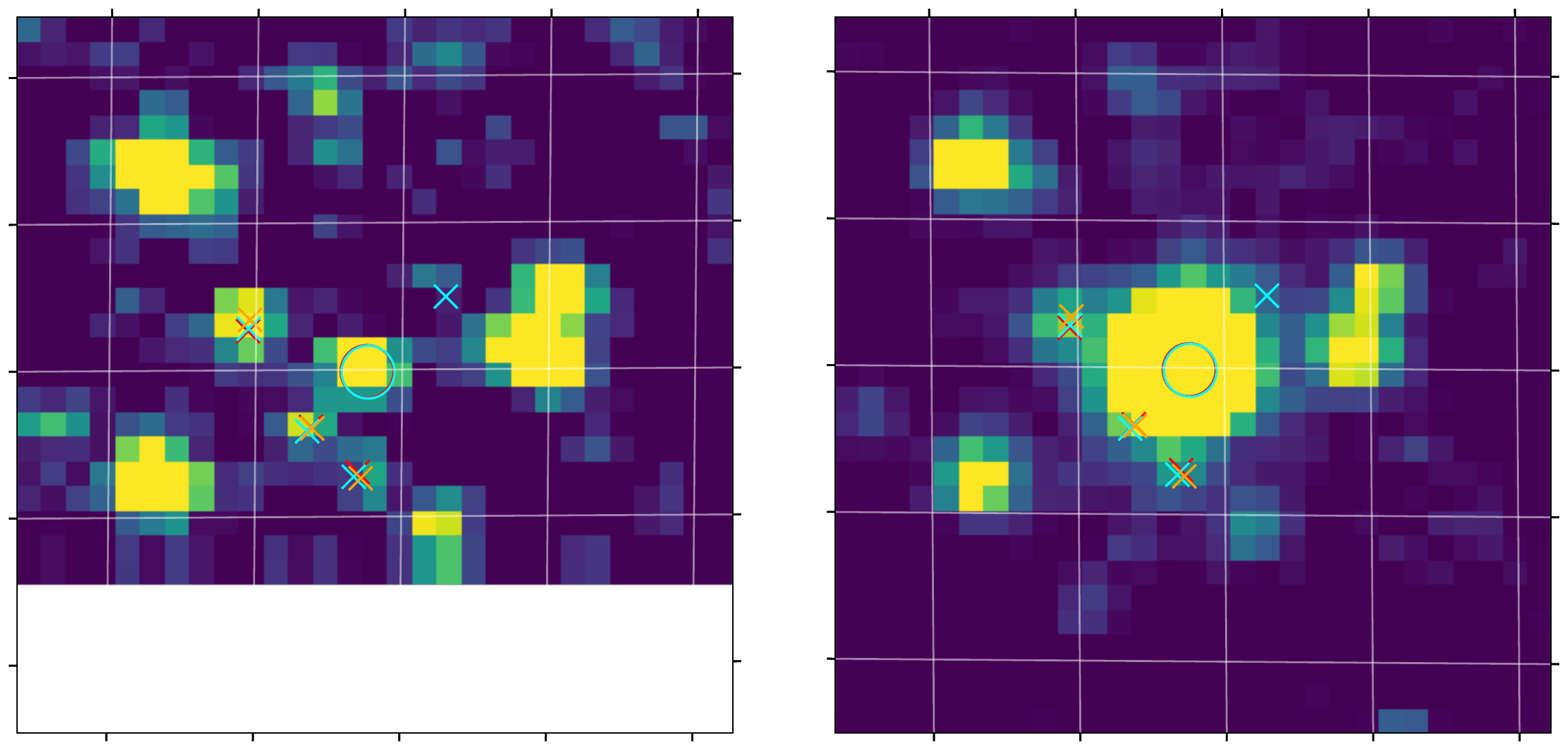}
    \caption{VVV \Ks{} band image cutouts from three days before minimum flux (left) and approximately two years later (right). Cutouts are $10$\arcsec{} on each side. The position of \wit{} is indicated by the circular marker, the positions of other sources within $2$\arcsec{} are indicated by crosses. VVV, DECaPS, and \gaia\ eDR3 positions are indicated in red, cyan and orange respectively. The white space on the left panel is due to the proximity of the target to the edge of the detector.
    }
    \label{fig:cutout}
  \end{center}
\end{figure}

The resultant $V$, $I$ and \Ks{} band light curves for \wit{} are shown in Figure \ref{fig:VIKs_lc}. \Ks{} band image cutouts demonstrating the high source density of the field are shown in Figure \ref{fig:cutout}.

\subsection{Survey photometry}

In addition to the $V$ and $I$ band optical and \Ks{} band near-infrared light curves from OGLE and \VVVx{} respectively, we acquired survey photometry from a number of additional sources, as presented in Table \ref{data}. Note that since the optimal extraction of photometry from OGLE and \VVVx{} images required extra care to address the nearby sources, so the target is likely to be blended to some degree in low angular resolution data such as GLIMPSE, longer wavelength WISE bands and possibly 2MASS.

\subsection{Spectroscopy}

\begin{figure*}
  \begin{center}
    \includegraphics[width=\textwidth,keepaspectratio]{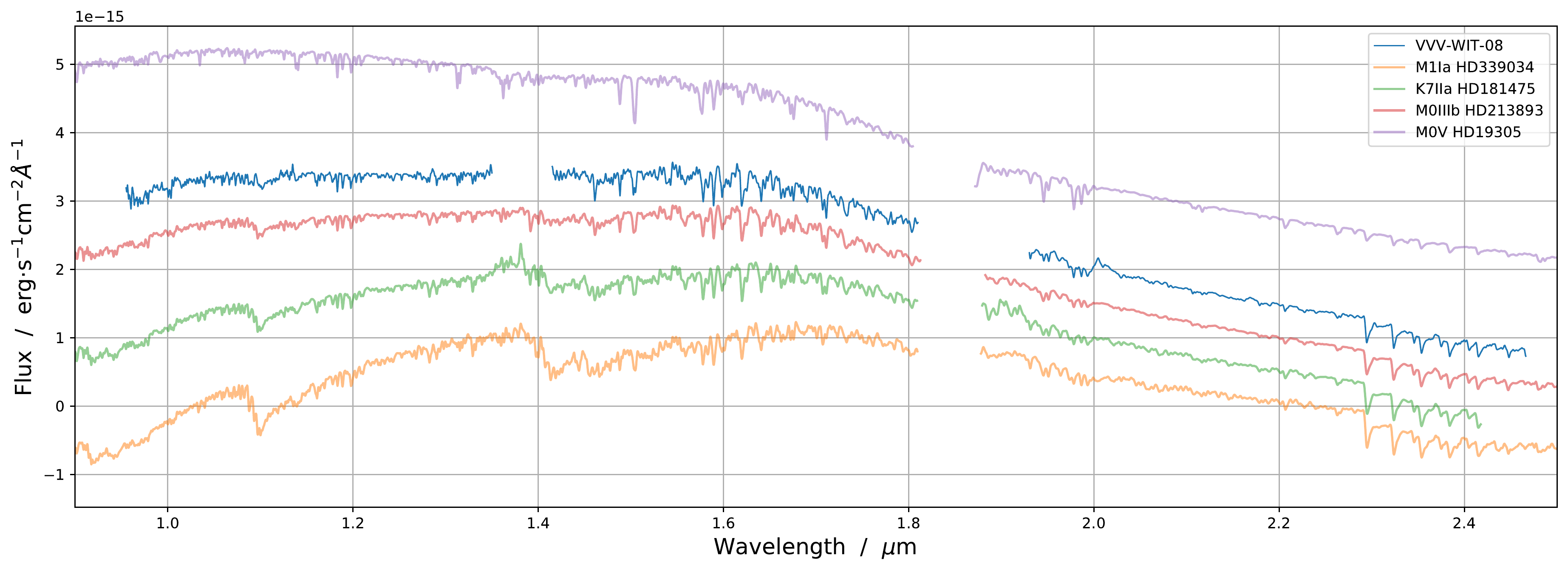}
    \caption{A TripleSpec near-infrared spectrum of \wit{} obtained on the 22 June 2019 (dark blue line). The spectrum is smoothed with a narrow ($\sim{}6.5$~\AA) Gaussian filter for clarity. In addition we include four template spectra of stars of a similar $T_{\rm eff}$ but different luminosity classes, obtained from the IRTF Spectral Library \citep{Rayner09}. The templates were normalised to the mean flux of \wit{} in the \Ks{} band, reddened using $A_V=2.17$~mag (see Section \ref{phot_analysis} for details), and then offset for clarity.}
    \label{fig:ts_soar_spec_08}
  \end{center}
\end{figure*}

Spectroscopic follow-up of \wit{} was performed at the 4-meter SOAR Telescope in Chile on June 22, 2019, during the Science Verification phase of TripleSpec 4.1 IR spectrograph \citep{2004SPIE.5492.1295W,2008SPIE.7014E..0XH}. The instrument covers simultaneously the wavelength range of 0.80\,$\mu$m to 2.47\,$\mu$m with resolving power of R$\sim$3500 and 1.1\arcsec{} wide slit. The collected spectra were reduced with a modified version of the IDL-based SpexTool \citep{2004PASP..116..362C} and flux calibrated against HD 162220. The resulting spectrum is presented in Figure \ref{fig:ts_soar_spec_08}.

\section{The Nature of VVV-WIT-08}\label{sec:physical_params}

\subsection{Spectroscopic analysis}\label{specanalysis}

A visual comparison of our near-infrared spectrum of \wit{} to those of the IRTF spectral library \citep{Rayner09} suggests that the star is a cool giant with a type no earlier than K7 and no later than M2, due to the general pseudo-continuum shape and the lack of FeH at 0.97\,$\mu$m respectively. Additionally, the CN band at 1.1\,$\mu$m is too weak to be consistent with a supergiant, as is evident in Figure \ref{fig:ts_soar_spec_08}.

A separate spectroscopic modelling by interpolated stellar templates from \citet{spectral_models} yielded a best fit effective temperature of $3614\pm{}3~K$, [M/H] of $-0.29\pm{}0.01$, $\log g$ of $1.05\pm{0.01}$, and a heliocentric radial velocity of $-14\pm{10}$~km~s$^{-1}$. The fitting was performed using a global optimization algorithm, implemented in the {\tt FER\reflectbox{R}E} code\footnote{{\tt FER\reflectbox{R}E} is available from \url{http://github.com/callendeprieto/ferre}} \citep{AllendePrieto06}. The provided uncertainties are statistical, but systematic uncertainties are likely significantly larger due to the relatively high signal-to-noise ratio of the near-infrared spectrum. Reasonable estimates for the systematic uncertainties based for a  typical cool giant are $50$~K, $0.5$ and $0.1$ for $T_{\rm eff}$, $\log g$, and [M/H] respectively.

\subsection{Photometric analysis}\label{phot_analysis}

\subsubsection{Isochrone fitting}\label{isochrone}

To estimate the physical properties of the \wit{}, we fit its spectral energy distribution (SED) using photometric data from various surveys presented in Table \ref{data} and Figure \ref{sed_model}.

We compare the photometric measurements with uncertainties presented in Table \ref{data} to MIST isochrone predictions \citep{mist0,mist1} shifted to various distances and with effects of interstellar extinction included.
In the modelling procedure we also take into account the $\log g$, $T_{\rm eff}$ and [M/H] constraints from the near-infrared spectrum, and \gaia\ eDR3 parallax measurements. To take into account possible systematic errors in isochrones/photometry/reddening we also include in the model an additional uncertainty (or excess noise) that is shared across surveys and is added in quadrature to each photometric measurement error.

The full model includes the following parameters: 1) Stellar age 2) Metallicity
3) Initial mass 4) Distance 5) $A_V$ extinction and 6) Excess photometric noise. The isochrone interpolation between metallicity, age, and initial mass is performed using \textsc{minimint}\footnote{\url{https://github.com/segasai/minimint}} code \citep{minimint}. 

To properly estimate the effects of extinction we use the ATLAS9 M0III $\log g=1.5$ solar metallicity model spectrum \citep{ck03} and the filter transmission from the SVO Filter Profile service \citep{SVOFPS} to estimate the effective wavelength of each filter for our target. These are then used to estimate $A_{\lambda{}}/A_{v}$ for the optical and near-infrared photometric bandpasses using equations 9 and 10 of \citet{Wang_Chen_19}. For the mid-infrared data equivalent relations are not provided, so we adopt the $A_{\lambda{}}/A_{v}$ values for a typical red clump giant provided in their table 3. These adopted mid-infrared values should be an adequate approximation for our target at these wavelengths due to their relative insensitivity to the optical extinction.

We adopt the following priors on model parameters. The stellar age is uniform between $10$\,Myr and $13$\,Gyr, metallicity is a Gaussian distribution with mean $-0.29$ and standard deviation $0.1$ (based on the spectroscopic constraints), extinction $A_V$ is uniform and within $0<A_V<4$. The prior on additional photometric scatter across different bands is log-uniform between $0.01$ and $0.3$ mag, and for mass we use the prior based on the \citet{Chabrier05} initial mass function with $M_{min}=0.5 M_\odot$ and $M_{max}= 10 M_\odot$.

To sample the posterior of the model described above we use parallel tempering ensemble sampling, implemented by the ptemcee package \citep{Vousden:2016,emcee}\footnote{\url{https://github.com/willvousden/ptemcee}}, with 1000 walkers and ladder of 5 temperatures that were run for 50000 steps after a burn-in of 50000 steps which were discarded.

{\renewcommand{\arraystretch}{1.5}
\begin{table}
\begin{center}
  \caption{Priors and median and 95th percentile credible intervals on the posterior pdf of the physical properties of \wit{} on requiring that ${\rm V}_{\rm tan}(R) < {\rm V}_{\rm esc}(R)$.}
\label{physical_params}
\resizebox{\columnwidth}{!}{
\begin{tabular}{|l|c|c|c|}
\hline
  \multicolumn{1}{|c|}{Parameter} &
  \multicolumn{1}{c|}{Prior} &
  \multicolumn{1}{c|}{Posterior} &
  \multicolumn{1}{c|}{Units} \\
\hline
[M/H]                & $\mathcal{G}(-0.29, 0.1)$        & $-0.18^{+0.16}_{-0.16}$         &  dex \\
Age                  & $\mathcal{U}(10^{-2}, 13.0)$     & $8.34^{+4.37}_{-5.08}$          &  Gyr \\
Mass                 & $\mathcal{IMF}(0.5, 10.0)^a$     & $1.0^{+0.4}_{-0.1}$             &  M\textsubscript{\(\odot\)} \\
A$_{V}$              & $\mathcal{U}(0, 4)$              & $1.67^{+0.28}_{-0.26}$          &  mag \\
Distance             & $\mathcal{LU}(1, 100)$           & $19.9^{+1.7}_{-3.7}$            &  kpc \\
$\sigma$             & $\mathcal{LU}(10^{-2},0.3)$      & $0.07^{+0.03}_{-0.02}$          &  mag \\
$T_{\rm eff}$        &                                  & $3659^{+75}_{-67}$              &  K \\
$\log g$             &                                  & $0.81^{+0.18}_{-0.13}$          &  dex \\
$\omega$             &                                  & $50^{+12}_{-4}$                 &  $\mu$as \\
Luminosity           &                                  & $0.7^{+0.2}_{-0.2}$             &  $10^3$~L\textsubscript{\(\odot\)} \\
Radius (r$_{\rm g}$) &                                  & $0.32^{+0.03}_{-0.06}$          &  au \\
Radius (r$_{\rm g}$) &                                  & $15.9^{+0.4}_{-0.4}$            &  $\mu$as \\
V$_{l}$              &                                  & $-493^{+140}_{-63}$             &  km~s$^{-1}$ \\
V$_{b}$              &                                  & $-15^{+12}_{-13}$               &  km~s$^{-1}$ \\
\hline
\end{tabular}
}
\end{center}
$^a$ \citet{Chabrier05} IMF prior\\
\end{table}
}

\begin{figure}
  \begin{center}
    \includegraphics[width=\linewidth,keepaspectratio]{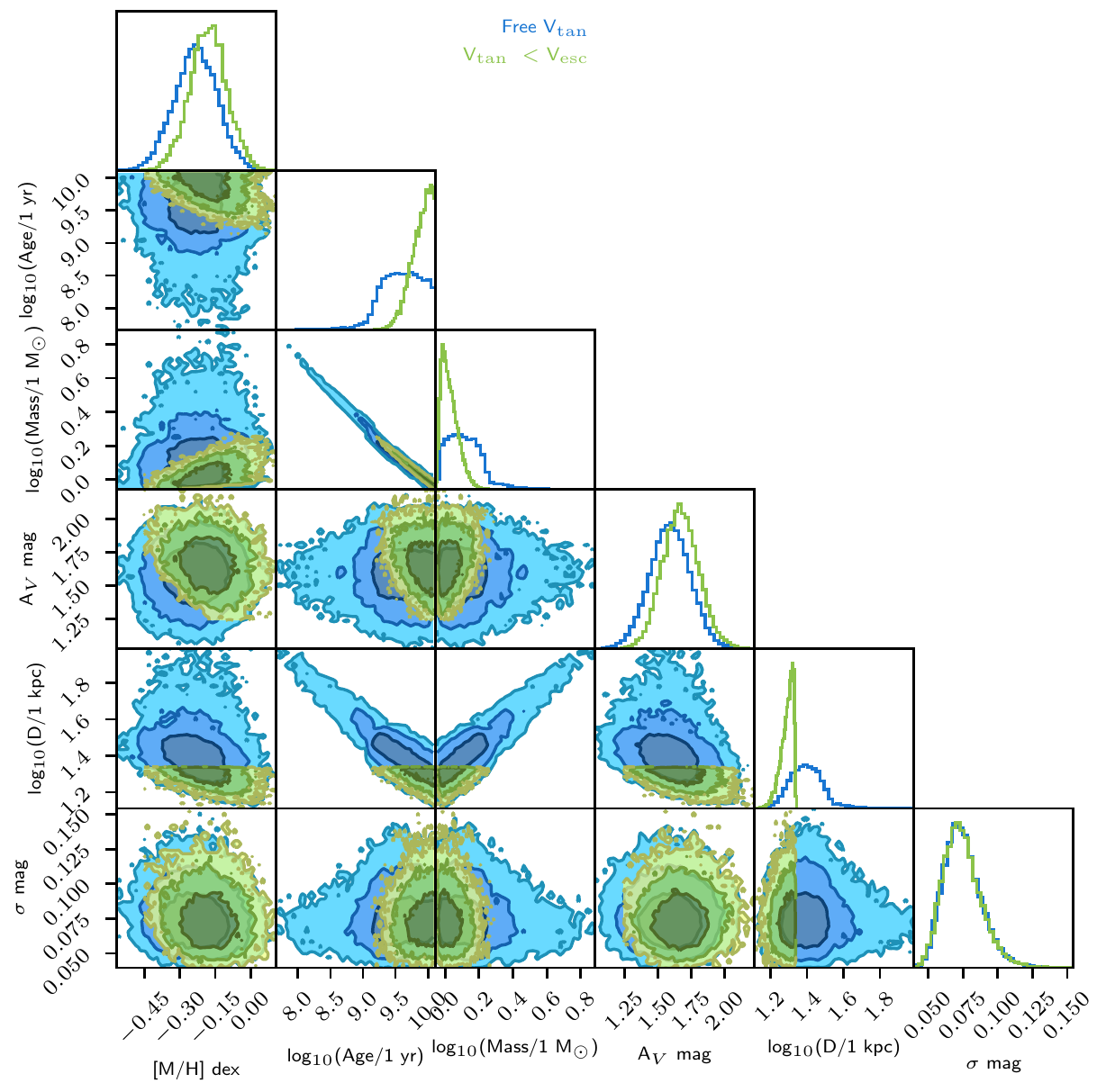}
    \caption{Posterior pdf on giant physical parameters obtained by modelling the photometry with MIST isochrones. The blue distribution is allowing an unrestricted tangential velocity, the green requires that the tangential velocity is lower than the Galactic escape velocity.}
    \label{giant_params_corner}
  \end{center}
\end{figure}

The preliminary isochrone modelling outcome, shown in Figure \ref{giant_params_corner}, suggests that the giant can be described by $1.0$ -- $2.4$ solar mass star at a range of distances from $\sim$ 17 to 42\, kpc at 2 $\sigma{}$. However, an important additional consideration is the range of allowable tangential velocities. It is extremely unlikely that \wit{} has a space velocity in excess of the Galactic escape velocity. If we require that ${\rm V}_{\rm tan}(R) < {\rm V}_{\rm esc}(R)$ then we can constrain the giant parameters further. ${\rm V}_{\rm esc}(R)$ is obtained using the \textsc{AGAMA} software package \citep{agama} with the \citet{mcmillan17} potential and the distance $(R)$ samples from the modelling. ${\rm V}_{\rm esc}$ is between $\sim{}500$ and $\sim{}600$~km~s$^{-1}$ at these distances and in this direction. We ignore the relatively small component of radial velocity for this estimate. Figure \ref{giant_params_corner} shows the impact of this requirement on the posterior probability distribution, and Table \ref{physical_params} summarises the inferred parameters and derived quantities. Essentially the part of the posterior with heliocentric distances larger than $
\sim$ $20$\,kpc is removed. These results suggest that the giant is likely a $0.9$ -- $1.4$ solar mass star at a heliocentric distance of between $16.1$ and $21.5$~kpc. The photometry of the model matches well the observations across all the bands requiring an additional systematic error of only $\sim0.07$~mag.
We adopt the ${\rm V}_{\rm tan}(R)<{\rm V}_{\rm esc}(R)$ requirement for all further analyses.

In the upper panel of Figure \ref{sed_model}, we show the best fit sample from the isochrone spectral energy distribution (SED) model versus the survey photometry. Also shown are the $68$ and $95$ per cent spread in the offsets relative to the photometric data points measured from a selection of random samples.

Crucially, from the above analysis we are able to obtain an estimate of the luminosity and hence radius of the giant. Evaluating the isochrone interpolant at each posterior sample, we obtain the median and $95$ per cent confidence interval on luminosity of $0.7^{+0.2}_{-0.2}$~$\times{}10^{3}$~L\textsubscript{\(\odot\)}. From the luminosity and $T_{\rm eff}$ samples, we obtain a median and $95$ per cent confidence interval on the stellar radius of $0.32^{+0.03}_{-0.06}$~au.

\begin{figure}
  \begin{center}
    \includegraphics[width=\linewidth,keepaspectratio]{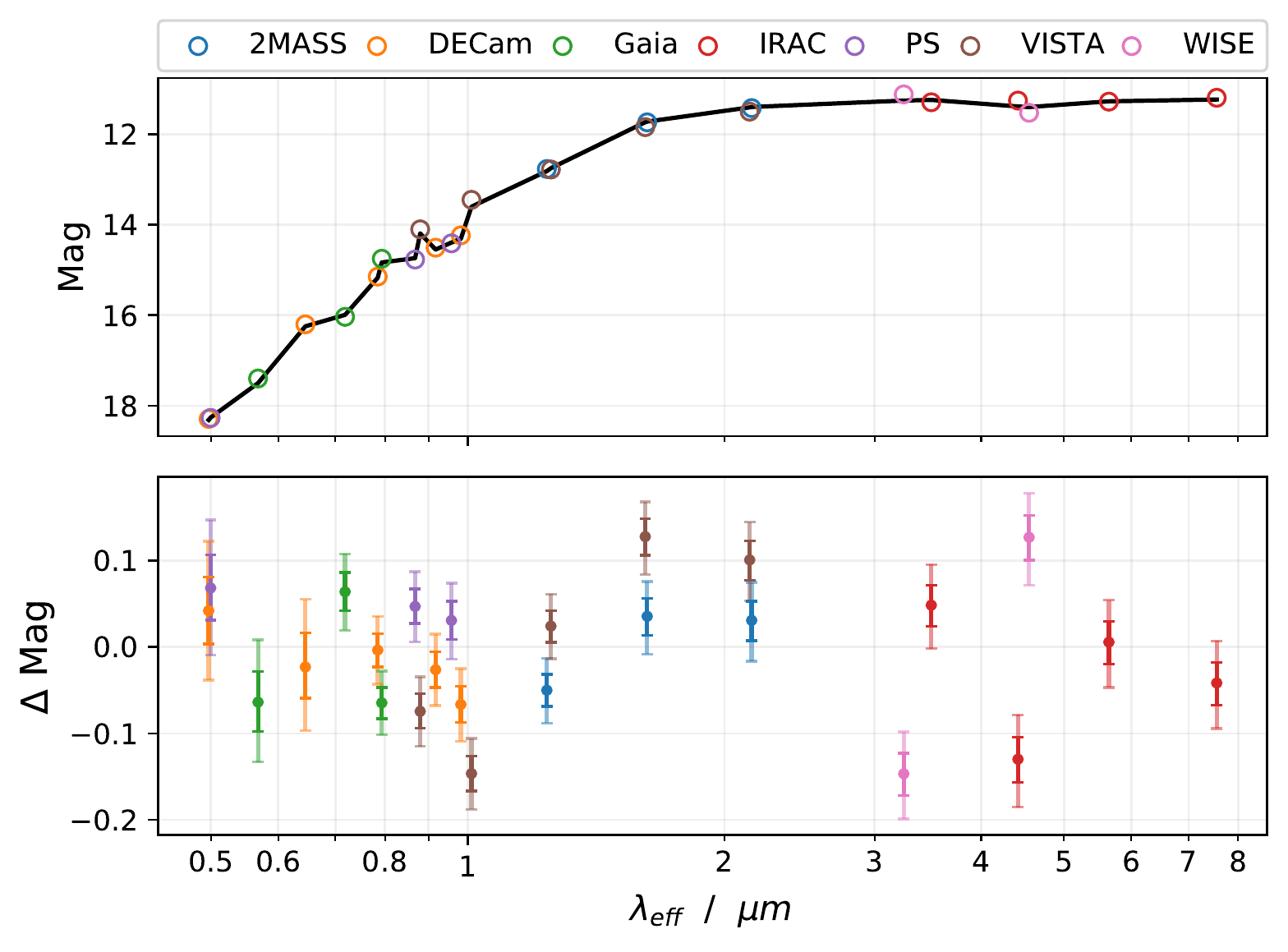}
    \caption{{\it Upper}:~Isochrone SEDs of a low $\chi^2$ sample from the posterior of the physical parameters of \wit{}. The black line shows a low $\chi^2$ model. {\it Lower}:~The offset from the median and spread in synthetic photometry, $95$ and $68$ per cent confidence intervals and the median are shown. The modelled excess error component $\sigma=0.07^{+0.03}_{-0.02}$~mag is not included.}
    \label{sed_model}
  \end{center}
\end{figure}

\subsubsection{Radius at Galactic bulge distance}\label{BB_Bulge_rad}

\begin{figure}
  \begin{center}
    \includegraphics[width=\linewidth,keepaspectratio]{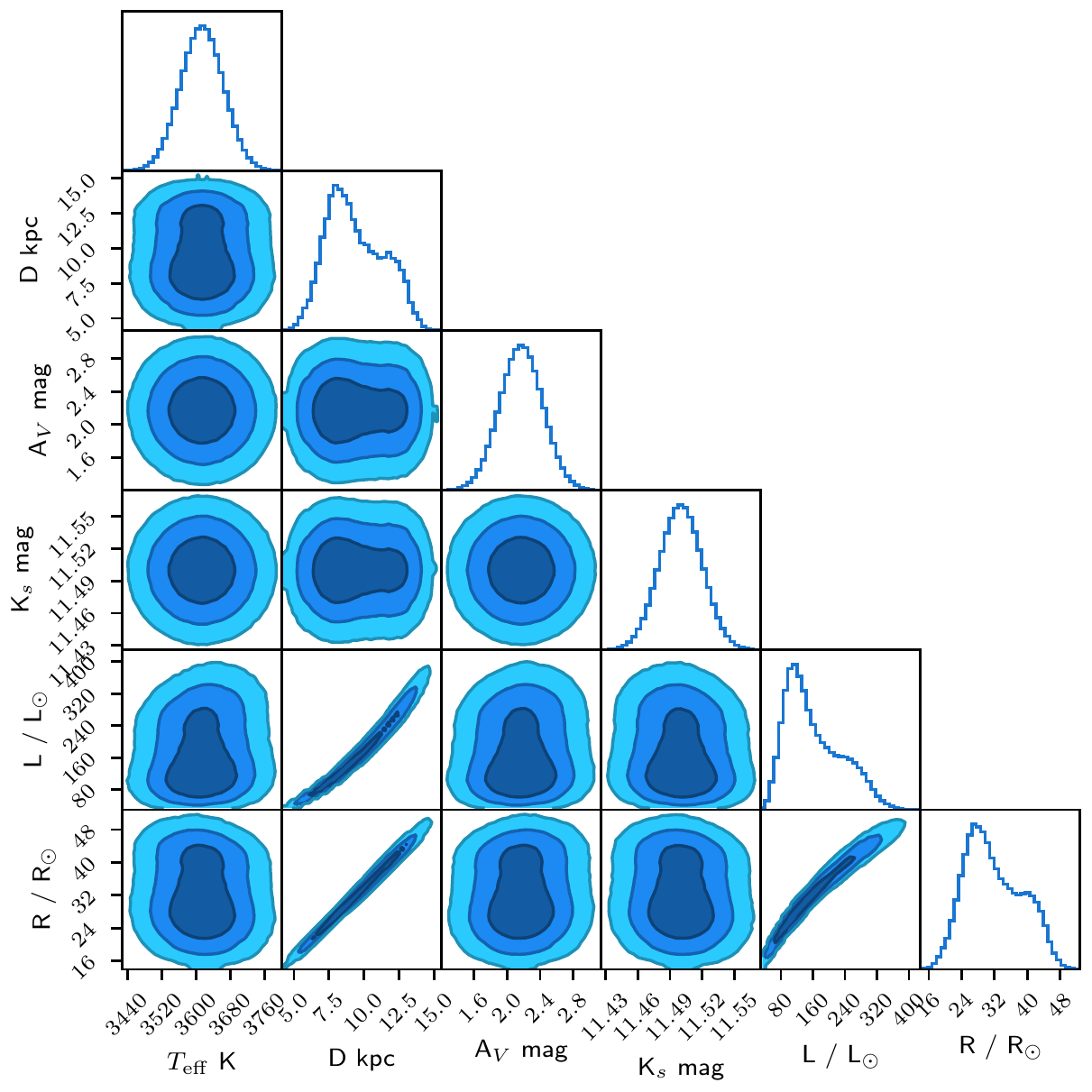}
    \caption{The pdf of the various independent and dependent parameters of the bulge distance physical parameter analysis. The distributions of $T_{\rm eff}$ and A$_V$ are from the analysis of the spectrum of \wit{} and nearby red clump stars respectively. The distance distribution is that of stars in a \texttt{Galaxia} model with similar position and proper motion to \wit{}. The \Ks{} band measurement assumes $0.02$~mag absolute calibration error. The distributions of luminosity and radius are extracted from $10^7$ random samples from the other parameters and model giant spectra.}
    \label{BB_corner}
  \end{center}
\end{figure}

Taking the \gaia\ eDR3 proper motion of the source and the isochrone fitting distance to the giant (with the ${\rm V}_{\rm tan}(R)<{\rm V}_{\rm esc}(R)$ requirement), we obtain tangential velocities of $-493^{+140}_{-63}$ and $-15^{+12}_{-13}$~km~s$^{-1}$ in the Galactic longitude and latitude directions respectively, having corrected for solar reflex motion. The lower bound on V$_{l}$ brings it closer to that of the Galactic rotation curve but it's still larger than might be expected. The distance distribution of stars from a \texttt{Galaxia} \citep{galaxia} model at the sky position of \wit{} that also share its proper motion within its uncertainties is shown in Figure \ref{BB_corner}. Note that a distance below $14.2~\mathrm{kpc}$ is ruled out to $3~\sigma$ from the isochrone analysis in Section \ref{isochrone} under the assumption that \wit{} is on an ordinary stellar evolutionary pathway. This is marginally consistent with the inference based on \texttt{Galaxia}.

The space velocity of \wit{} implied by the analysis in Section \ref{isochrone} appears halo-like, though this is in tension with its relatively metal rich spectrum ([M/H]$=-0.29$). We note that ${\rm V}_{b}$ and ${\rm V}_{los}$ are more moderate. These velocity components and the metallicity appear more consistent with a disc or bulge membership.

Since the proper motion of the star from VVV and \VVVx{} astrometry based on \gaia\ DR2 stars as astrometric reference points matched well the one measured by \gaia\ within 2$\sigma$ and therefore is likely accurate, we consider that the physical parameters of the giant star obtained through fitting of the isochrones to the photometry are relatively unlikely {\it prima facie}. We note that in addition to the metallicity, the proper motion is consistent with those of Galactic bulge stars.

To assess what distance would be expected from this proper motion alone we use the \texttt{Galaxia} simulation at the coordinates of \wit{} described above, having selected stars that share its proper motion on taking into account the \gaia\ eDR3 uncertainties. Other than a tiny component of nearby disc stars, which we remove, there are two modes in the distance distribution corresponding to stars in the bulge and the far side of the disc. The heliocentric distance range of $6.0$ to $13.3$~kpc contains $95$ per cent of the \texttt{Galaxia} sample.

The difference between the kinematic and isochrone distances may be interpreted as as lower-than-expected luminosity of \wit{} than predicted by isochrones and can perhaps be explained if we postulate that it did not follow a typical evolutionary pathway, as a fit to the MIST isochrones would require. There are examples in the literature of under-luminous giant stars, namely sub-subgiant stars or SSGs~\citep{otoossgs1,otoossgs2,otoossgs3}. These are redder than ordinary main sequence stars and fainter than giant and subgiant stars. SSGs are thought to form through binary evolution channels, although other formation mechanisms have been proposed \citep{otoossgs3}.
The majority of SSGs are variable to a significant degree, either in their photometry or radial velocity. While \wit{} is not significantly variable in the time series photometry we have (see Section \ref{quiescent_var}) -- barring the dipping event itself -- radial velocity follow up would be worthwhile. Approximately one third of SSGs are $H\alpha$ emitters \citep{otoossgs1}, but this does not appear to be the case for \wit{} in the VPHAS$+$ photometry.

The modelling of exotic stellar evolutionary channels is difficult. An attempt to find suitable parameters using BPASS binary evolution models \citep{bpass} could not be reconciled with the ordinary atmospheric parameters obtained from the spectrum of \wit{} and hence was fruitless. With this caveat in mind, we attempted to estimate the radius of \wit{} by considering simply the luminosity required to obtain the measured \Ks{} band magnitude given the $T_{\rm eff}$, $\log g$, and [M/H] parameters obtained from the NIR spectrum and the distance pdf obtained from the \texttt{Galaxia} model as described above.
We include an additional $0.02$~mag uncertainty to take into account the typical absolute calibration error. The \Ks{} flux measurement should be the least impacted by uncertain extinction without resorting to using the mid-infrared photometry which is likely to suffer from blending. Through a measurement of $E(J-K_s) = 0.36\pm0.04$~mag for nearby red clump stars using VISTA PSF photometry and the \citet{Wang_Chen_19} law we obtain an estimate of $A_{V} = 2.17\pm{}0.26$~mag for this field, which we adopt. We used appropriate ATLAS model spectra \citep{ck03} and the synphot python package \citep{synphot} to perform synthetic observations and estimate stellar luminosities and hence radii. Sampling from the probability distributions of these parameters $10^7$ times yields a radius distribution of $0.14^{+0.07}_{-0.05}$~au ($31^{+14}_{-10}$~R\textsubscript{\(\odot\)}, median and $95$ per cent confidence interval).
Figure \ref{BB_corner} shows the probability density functions of all parameters.

The measurements made assuming a Galactic bulge location will be retained in addition to the isochronal measurements for the purposes of further analysis. We consider the former to be sensible lower limits on the distance and (more importantly) radius of the giant, where the latter are sensible upper limits. It's highly likely that the distance and radius of \wit{} lies somewhere in between these two scenarios.

We note that if we adopt a Galactic bulge location, an alternative explanation for the low luminosity of the giant would be that there is diffuse matter in the star system or along the line of sight that is causing grey extinction. We have no reason to think this is correct but given the potential presence of optically thick material causing a large drop in flux it cannot be ruled out. In this scenario it would be appropriate to adopt the isochronal radius distribution, which is the most important parameter for the remainder of this investigation.

\subsection{Quiescent variability}\label{quiescent_var}

Close inspection of the light curve presented in Figure \ref{fig:VIKs_lc} reveals a quasi-periodic signal in the $I$ band data outside the event. We also find it is necessary to inflate the OGLE $I$ and $V$ band photometric uncertainties through the addition in quadrature of an excess error \mbox{$\sigma_{{\rm excess},V,I}=0.019, 0.012$ mag} in order to explain the scatter in their extra-event light curves. Uncertainties in the $K_s$ band data for this source appear to be overestimated by approximately $40$ per cent at baseline flux. We suspect DoPHOT is overestimating the uncertainties near saturation that occurs close to $K_s=11.5$~mag in VVV and \VVVx{} observations with good seeing.

It remains possible that this apparent variability could be explained by residual systematic errors in the data, but we consider this less likely given the care taken when reducing these data. We also note that the $V$ and $I$ band scatter is mildly correlated.

To attempt to measure a period in the variability while taking into account possible time-varying stellar surface conditions, we produce mean Lomb-Scargle periodograms across moving windows in the extra-event $I$ band data using a sequence of different window widths. We find a weak period of approximately $195$~days (and its alias at approximately $390$~days) and a weaker one at approximately $124$~days. If genuine, these periods would be roughly consistent with rotation speeds of red giants measured with {\it Kepler}, in particular a $124$~day period would fit nicely within the distribution shown in figure 5 of \citet{Ceillier17}. The magnitude of the excess scatter appears to be consistent with that of ordinary variability of some red giant stars as measured by \citet{Henry00}, and \citet{Olah18} found that approximately half of giants exhibit flux variability above $1$ per cent.

A \lq back of the envelope' estimate of the flux variability expected from a single star spot gives around $1.8$ per cent in the $V$ band and $1.3$ per cent in the $I$ band, assuming black body spectra and requiring $\Delta{}T=200$~K with a spot covering $5$ per cent of the giants surface. For reference, the variability in the \Ks{} band using this model is $0.6$ per cent.

It is worth bearing in mind that the above tentative period of the quiescent variability is relatively close to the duration of the major dimming event. This should be taken under consideration during analysis of the event itself. In particular we suspect that the subtle difference in the flux at baseline immediately before and after the event is due to this quiescent variability.

\section{Event analysis}\label{event_modelling}

\subsection{Intrinsic variability}

We are not aware of any mechanism of intrinsic stellar variability which could cause the observed drop in flux of this magnitude across the wavelength range $0.5-2.2\mu{}$m. The shape of the dipping event and relative constancy of its colour are additional strikes against the intrinsic variability hypothesis. We briefly discuss two specific possibilities, before discarding them.

\subsubsection{R Coronae Borealis-like variability}

R Coronae Borealis (RCB hereafter) are rare, hydrogen-deficient carbon supergiant stars. Their light curves are characterised by deep (up to $9$ mags in $V$ band) reductions in their luminosity due to the condensation of carbon to soot in their atmospheres. RCB stars have warm shells with \mbox{$300<T_{\rm eff}~(K)<1000$} and hence tend to exhibit a mid-infrared excess \citep{Tisserand12}, whereas this does not appear to be the true of \wit{}. Major dimming events of RCB stars are characterised by a sharp ingress and a relatively shallow egress. In principle, the \wit{} event light curve is consistent with this due to the gap in the ingress data but its relative achromaticity is not -- RCB dimming events are significantly deeper at shorter wavelengths \citep{Feast97}.

\subsubsection{Variable young stellar object}

High amplitude photometric variability is common in YSOs, often caused by episodic accretion. Dips are also known to occur. Low amplitude dips typically have timescales from hours to days \citep{Findeisen13,Cody14,Scaringi16, Ansdell16} but deeper dips on timescales of years are also seen \citep[e.g.][]{Bouvier13,cp17a}, attributed to a substantial perturbation of the accretion disc structure, perhaps by a young planet.
However, \wit{} does not appear to reside in or near a region of the sky with obvious signs of ongoing star formation. Its spectrum shows no signs of youth, and its lack of significant mid-infrared excess indicates similarly. Outside the event, \wit{} has low variability and the dimming event itself is largely achromatic, both of which are uncharacteristic of YSO variability.

\subsection{Occultation}

As we are unable to explain the behaviour of \wit{} through intrinsic variability, we turn our attention to occultation.

\subsubsection{Preliminary considerations}

If an occulter is orbiting the giant, then the long OGLE $I$ band time baseline rules out an orbital period shorter than around $11$~years.

The time baseline between the first point of reduced flux and the approximate last point of reduced flux, and hence the minimum event width, is around $170$~days. Additionally, in this sight line and due to the distance to the giant, some portion of the Galactic bulge is in the foreground. The long timescale of the event, and the possibility that the occulter is at an intermediate distance means that relative parallax between the giant star and the occulter may need to be considered.

As might be expected given its position and as is demonstrated in Figure \ref{fig:cutout}, the target is subject to significant crowding. The nearest on-sky objects are $1.06$\arcsec{} and $1.44$\arcsec{} away in the \gaia\ DR2 data, and there is an additional source at $1.74$\arcsec{} separation in the VVV and DECaPS data which was not detected by \gaia. None of these neighbours exhibit clear dimming events similar to that of \wit{} in our data. This means any occulter must either be more distant than the nearby objects, or have angular size in the direction of the nearby stars that is smaller than their separations and a motion vector which does not pass in front of them during our observational coverage.

The lack of obvious amplification in the light curve due to microlensing implies that the angular size of an occulter must be much larger than its Einstein radius.

There is a several hundred day gap in our observational coverage during most of the 2009 observing season. In principle, a similar event could have occurred in this year and been missed, but a periodic signal of $\approx{}3$ years is ruled out due to the lack of similar events in 2003, 2006, and 2015.

Due to surface brightness considerations, the depth of the event rules out a common eclipsing binary system with the second component being another star. To produce a 97 per cent reduction in the $I$ band flux the secondary would need to have an equally large radius and an effective temperature below 1400\,K. The only remaining binary system parameter space is a large ($r\gtrsim r_{\rm g}$) and dark or self obscured companion.

\subsubsection{Chance alignment probability}\label{space_density}

If the occultation is due to the presence of an object unbound from the background star at some intermediate distance then we are able to estimate the implied space density required in order to have a reasonable chance of our finding \wit{}.
On adopting the approximation that occulters and background stars are circular in projection and ignoring the effects of parallax motion, if we have $\Sigma_o$ surface density of occulters and $\Sigma_s$ surface density of background stars, moving with relative velocity $V$ with respect to occulters, then the rate of encounters with minimum separation between $r$ and $r+\mathrm{d}r$  per unit time $\mathrm{d}t$  occurring in the area $A$ will be
$$\mathrm{d}R= 2 V \Sigma_o \Sigma_s A \mathrm{d}r \mathrm{d}t.$$
Occultation duration and depth are functions of $V$ and the radii of the occulter and background star ($r_o$, $r_s$). We consider depth purely as a fraction of the background stellar disc that is obscured, ignoring limb darkening for simplicity, and ignoring the potential for flux contamination or a semi-opaque occulter.

An estimation of rates requires knowledge of both the stellar and occulter density and the relative velocity distribution, and for these we employ a Galactic population model. We ran a \texttt{Galaxia} simulation at the location of \wit{}, selecting objects with no regard to apparent or absolute magnitude as potential occulters, and those with $K_s<17$ (which is the approximate limit of our VVV psf data) as potential background sources. Using these selections, and the further requirement that the background star is more distant than the occulter, we evaluate the rates of occultations of different durations ($p$) and depths ($\Delta{f}$; depth of one corresponds to full occultation). Each sample uses the \texttt{Galaxia} distances and velocities for both components and for the radius of the background star, but the occulter radius we fix at $0.5$ au.

The rates obtained for $p>10$ day occultations are $0.80$ and $0.73$ events per square degree per year for $\Delta{f}>0.5$ and $\Delta{f}>0.9$, respectively. For $p>100$ day events these rates drop to $1.3\times{}10^{-2}$ and $1.2\times{}10^{-2}$ ${\rm deg}^{-2}{\rm yr}^{-1}$ for the same $\Delta{f}$ limits. We note that there is significant volume in the distribution at $\Delta{f}=1.0$, which is to be expected where $r_o\gg r_s$. If we could be certain that the remaining 3 per cent flux seen in the \wit{} event were due solely to an incomplete obscuration then we might additionally incorporate the requirement that $\Delta{f}<1.0$. In this case, for $p>100$ day events with $0.9<\Delta{f}<1.0$ the rate is $3.4\times{}10^{-4}~{\rm deg}^{-2}{\rm yr}^{-1}$.

We consider the following additional modifiers to the above rates. The \texttt{Galaxia} simulation predicts a $K_s<17$~mag source density of $10^6$ stars per square degree and the rates scale proportionally relative to this. The rates reported are under the assumption that the occulter space density is equal to the stellar space density. Changing the adopted occulter space density will alter the rates proportionally. Provided that $r_o\gg r_s$, the rate of all occultations can be directly scaled up or down to take into account larger or smaller occulters, although the depth and duration distributions will change. $0.5$~au is appropriate for \wit{} (see Table \ref{tab:occulter_params}).

Ignoring the $b<-5\ {\rm deg}$ region, where the source density is much lower, the VVV survey bulge component is $\sim{}200\ {\rm deg}^{-2}$ and covers 10 years of observation. These 10 years of observation are punctuated by $\sim{}$ 6 month seasonal gaps in observation, reducing detection efficiency by $\sim{}50$ per cent. In our VVV psf data we find a source density that is approximately $50$ per cent larger than \texttt{Galaxia}'s prediction of $10^6\ {\rm deg}^{-2}$. Using the $1.2\times{}10^{-2}$ ${\rm deg}^{-2}{\rm yr}^{-1}$ rate from above for $p>100$~day, $\Delta{f}>0.9$ events, we should expect to find $18$ \wit{}-like events in our data, with the important caveat that this assumes the space density of occulters is equal to stars. An additional caveat is that the rates above are appropriate for background stars that are brighter than $K_s=17$~mag, the approximate detection limit of our data, but stars must in fact be brighter than this at baseline such that we can identify a reduction in brightness. The exact baseline magnitude at which we are sensitive to dips is difficult to determine, but is likely to be a magnitude or two brighter than the $K_s=17$~mag detection limit of the survey. As a result the expected event rate will be a factor of a few lower, and hence in order to expect to find \wit{} the required occulter space density must to first order be similar to that of the stellar content.

While in principle it is possible that a population of dark occulting bodies exists that is similar in number to the stellar content -- if so it would take a concerted search to find them (that to our knowledge has not been undertaken in earnest), as we have just demonstrated -- we consider it unlikely. Furthermore, \wit{} is something of an anomaly itself (in either velocity or radius, see Sections \ref{isochrone} and \ref{BB_Bulge_rad}), and a rare event occurring to a rare object compounds the unlikeliness. As such, we deem it improbable that a chance alignment of an unrelated object is the cause of the dimming event. We adopt an occulter bound to the giant as our primary hypothesis.

\subsubsection{Modelling}\label{modelling}

A plausible explanation of a smooth, achromatic dip in flux is an occultation by an optically thick body. We first modelled the light curve as a solid spherical occulter, passing on a straight line relative to the background star. This modelling was implemented using the publicly available \textsc{batman} stellar transit light curve modelling software \citep{batman} and the \textsc{dynesty} dynamic nested sampling package \citep{dynest, dynesty}, and yielded a mediocre fit to the light curve with $r_{\rm o}/r_{\rm g}\approx{}1.3$. The best fit we were able to achieve is shown in Figure \ref{occultation_models} (red line), and demonstrates that a spherical occulter on a straight line trajectory cannot adequately match the data at hand.

This led us to enhance the model by allowing the occulter to be elliptical with finite transparency and to have a parallax with respect to the giant star. This is partly motivated by \citet{Rappaport19}, who showed that an occultation by an object that is elliptical in projection is able to explain a smooth, asymmetric light curve in the case of EPIC~204376071.

We implemented the light curve generation ourselves using custom written \code{C++} software that we release with the paper\footnote{\url{https://github.com/segasai/vvv_dipper_code}}. Specifically, this software integrates over the obscured part of the disc of the star, taking into account the limb darkening. The integration is done using a pre-calculated grid of 1000$\times$1000 points on the disc of the star. The \code{C++} software is optimized for maximum performance and is parallelized using \code{OpenMP}.

Analysis of the space density of dark occulters required in order to produce a \wit{}-like event (Section \ref{space_density}) indicates that there is no need to include parallactic motion in the model. Despite this, we were curious to see whether there was volume in the posterior for high parallax models. Considerable effort was expended in trying to extract the posterior pdf of the model parameters with parallax included. We attempted the modelling with a large selection of currently available nested and ensemble sampling packages. Despite intensive effort, careful reparametrisation to reduce degeneracies, and in some cases modification of the sampling packages themselves, none were capable of satisfactorily capturing the complex shape of the posterior and usually missed one or more known modes. This was principally due to the narrow likelihood surface, multiple clearly distinct modes, and vast regions of the posterior with a flat likelihood.

While we know models containing a large relative occulter parallax exist that are capable of explaining our data, the analysis in Section \ref{space_density} indicates these are unlikely a priori. Due to the need to further reduce degeneracies in order to produce a satisfactory posterior we ultimately chose to remove this parameter from our model. In doing so we reduce the motion vector to a straight line relative to the giant star, which can be parametrised as a scalar value, reducing the velocity posterior distribution to a single mode and removing the velocity direction parameter.

The final model that we implement has the following parameters and priors:

\begin{enumerate}
\item The semimajor axis $r_{\rm maj}$ of the elliptical occulter. It is measured with respect to the angular radius of the star $r_{\rm g}$. The range over which we sample is $-0.5<\log_{10} r_{\rm maj} < 2$  with the prior $r_{\rm maj}^{-2}$. We discuss the need for this prior below.
 
\item The semiminor axis $r_{\rm min}$ of the elliptical occulter. Again measured with respect to the angular radius of the star $r_{\rm g}$. The prior range is $-0.5<\log_{10} r_{\rm min}<2$ with an informative Beta prior on the axis ratio $r_{\rm min}/r_{\rm maj}=q \sim {\mathcal B}(1, 0.5)$.

\item The orientation angle of the occulter major axis $\theta$, which is measured with respect to the velocity vector. We use the prior $\theta \sim \mathcal{U}(0, \pi)$.

\item The impact parameter (radius) $r_{0}$ of the occulter relative to the giant at time $t_{0}$, i.e. the minimum separation between the occulter and giant measured in units of the radius of the star $r_{\rm g}$. The prior is $\log_{10} r_{0} \sim {\mathcal U}(-3, 2)$.

\item The tangential proper motion vector of the occulter relative to the giant, $\mu$. The proper motion is measured in units of radius of the giant per day $\mu \sim \mathcal{U}(0.0, 0.2)$.

\item The quadratic limb darkening parameters of the giant $a$ and $b$, defined so that
the specific intensity on the surface of the star $I(\mu)=1 - a (1-\mu) - b (1-\mu)^2$, where $\mu = \cos \Theta$ and $\Theta$ is the angle between the line of sight and the normal to the star surface.
$a\sim {\mathcal U}(0.5,0.7)$, $b\sim {\mathcal U}(0.05,0.25)$. Bounds on the quadratic limb darkening parameters $a$ and $b$ cover the range of values expected for stars with similar $T_{\rm eff}$, $\log g$, and [M/H] to \wit{} in the $I$ band as computed by \citet{ld_bounds}.

\item The time $t_{0}$ at which the position of the occulter is $x_0,y_0$. It is the time of the minimum on-sky separation between the occulter and giant. The prior is $t_0 \sim \mathcal{U}(54500, 57500)$

\item The transparency of the occulter $T$, where values of 0 or 1 correspond to a completely opaque or transparent occulter respectively and the prior is $\log_{10} T\sim \mathcal{U}(-5,0)$.

\end{enumerate}
The model parameters, priors and posteriors from the modelling are also summarized in Table \ref{model_params}.

Two aspects of the prior require a special remarks. The reason for the $r_{\rm maj}^{-2}$ prior was to correct for the fact that large occulters are intrinsically more likely to produce occultation events due to their large cross-section. With the $r_{\rm maj}^{-2}$ prior our posterior is characterizing the information we gained on the occultor conditional on the event taking place. The prior on the axis ratio $r_{\rm min}/r_{\rm maj}=q \sim {\mathcal B}(1, 0.5)$ was introduced to deal with a degeneracy that our data is mostly constraining the semiminor axis of the occulter with essentially no constraints on $r_{\rm maj}$,  thus allowing arbitrarily elongated objects. The prior on the axis ratio penalizes such solutions.
This prior has $\sim$ 31 per cent of prior volume in the $q>0.9$ region and 39 per cent prior volume for $0.5<q<0.9$.  Since the prior is informative and is affecting the posterior, caution must be exercised when interpreting the inference of $r_{\rm maj}$ and $r_{\rm min}$.

We parametrise the occulter as having transparency between negligible and total. Note however that transparency is wholly degenerate with flux contamination, the contaminant being the occulter or an unrelated background or foreground star. In the absence of other information, we are unable to differentiate between flux contamination and occulter transparency.

The model specified above could in principle be applied to a data set with photometric measurements in multiple bands while requiring different limb darkening parameters ($a, b$) for each but since we have an order of magnitude more data points in the $I$ band than in the $V$ and \Ks{} bands, we chose to model the $I$ band data only.
The model makes a prediction for the fraction of flux of the star as a function of time $f(t)$. We then use those predictions to compare with the observed fraction of the star's flux with respect to the baseline value of $I=14.713$~mag. To deal with the fact that the light curve of the star outside the event shows a slight flux variability (caused by either uncorrected systematics or low-level intrinsic variability, see Section \ref{quiescent_var}), we inflate the magnitude uncertainties by $0.012$~mag: $\sigma_{infl}= \sqrt{\sigma_0^2  + 0.012^2}$.

\begin{figure*}
  \begin{center}
    \includegraphics[width=\linewidth,keepaspectratio]{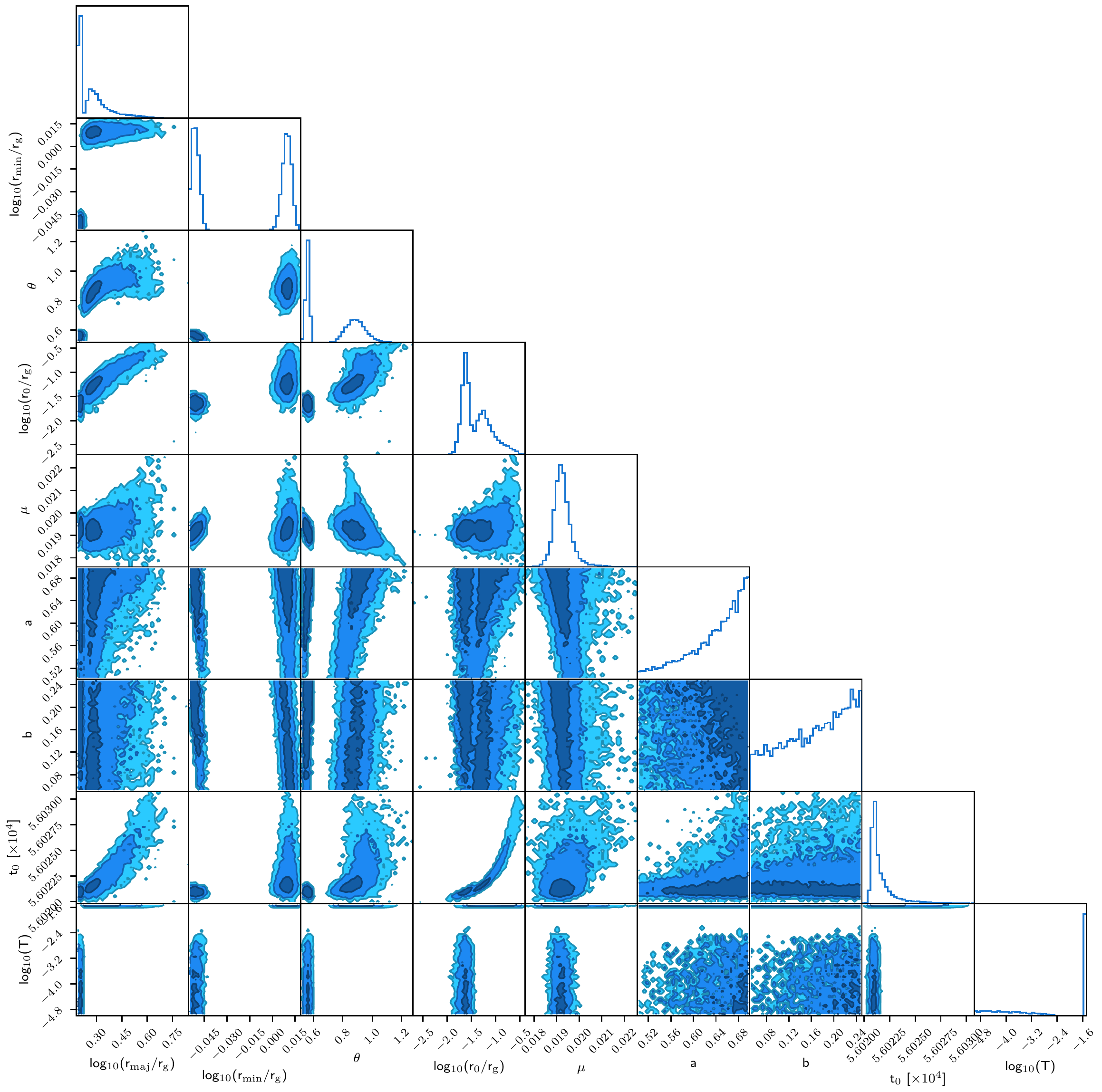}
    \caption{Posterior pdf on elliptical occultation model parameters. Length parameters (occulter size, position and motion) are in units of giant radii ($r_{\rm g}$) and motions are relative to the giant. Times are in units of modified Julian days and $\theta$ is in units of radians. The contours correspond to $1\sigma$, $2\sigma$ and $3\sigma$ limits. The median and $95$ per cent confidence interval on these parameters are given in Table \ref{model_params}}
    \label{fig:occultation_corner}
  \end{center}
\end{figure*}

\begin{figure*}
  \begin{center}
    \includegraphics[width=\linewidth,keepaspectratio]{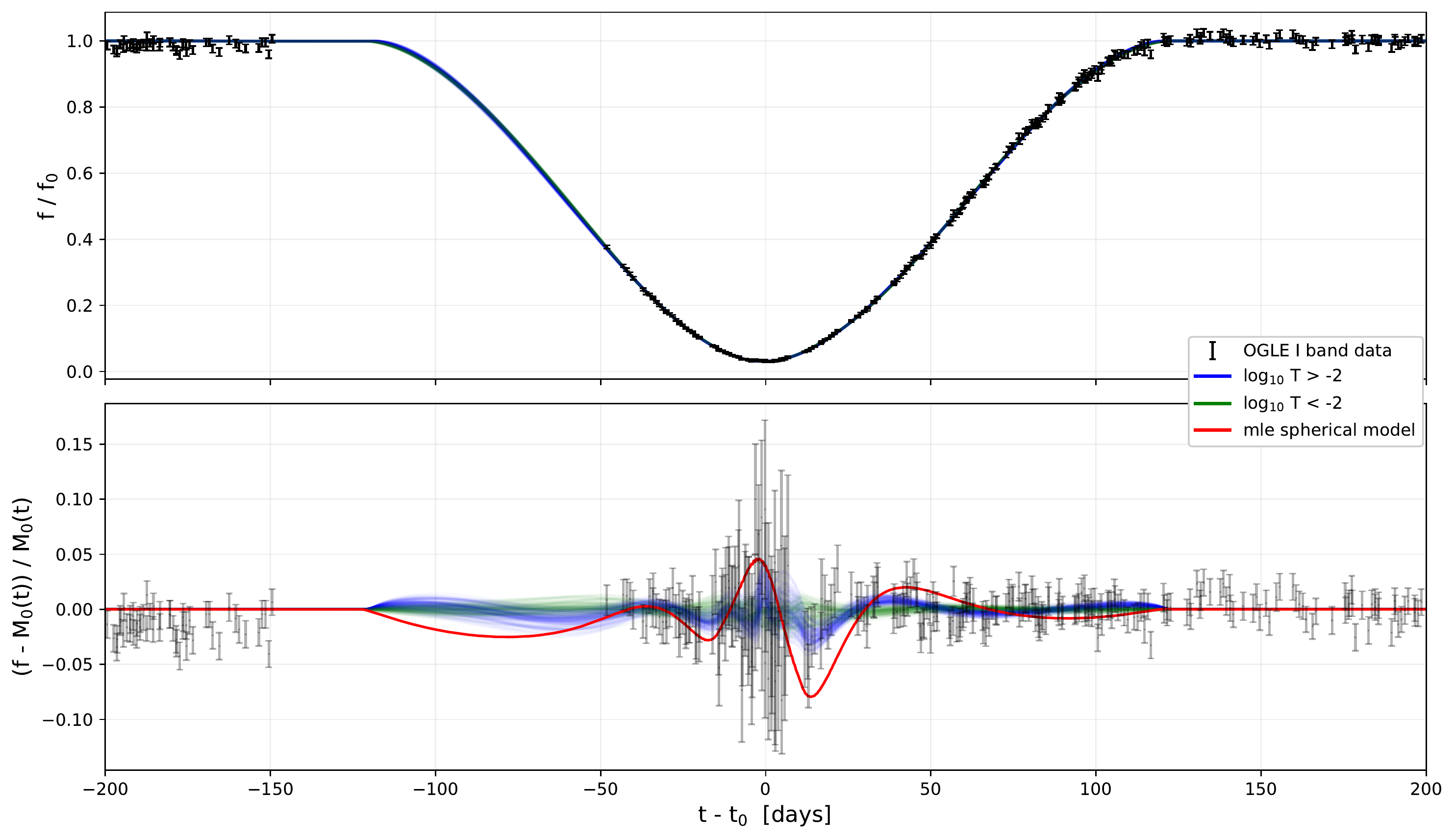}
    \caption{\textit{Upper:} A random selection of $200$ light curve models from the posterior pdf with the OGLE $I$ band measurements from which they were obtained. \textit{Lower:} Fractional residuals to the maximum likelihood estimate (MLE) model. The maximum event depth in the $I$ band as measured from these models is $96.93^{+0.05}_{-0.06}$ percent at $1\sigma$. In both panels the light curve models are rendered in blue where they correspond to the high transparency mode, and in green where they correspond to the low transparency mode. The error bars do not reflect the additional $0.012$~mag uncertainty added in quadrature to take into account quiescent variability. The lower panel includes in red the difference between the elliptical occulter maximum likelihood model and the best fit we were able to achieve with an occulter that is opaque and circular in projection.}
    \label{occultation_models}
  \end{center}
\end{figure*}

\begin{figure*}
  \begin{center}
    \includegraphics[width=\linewidth,keepaspectratio]{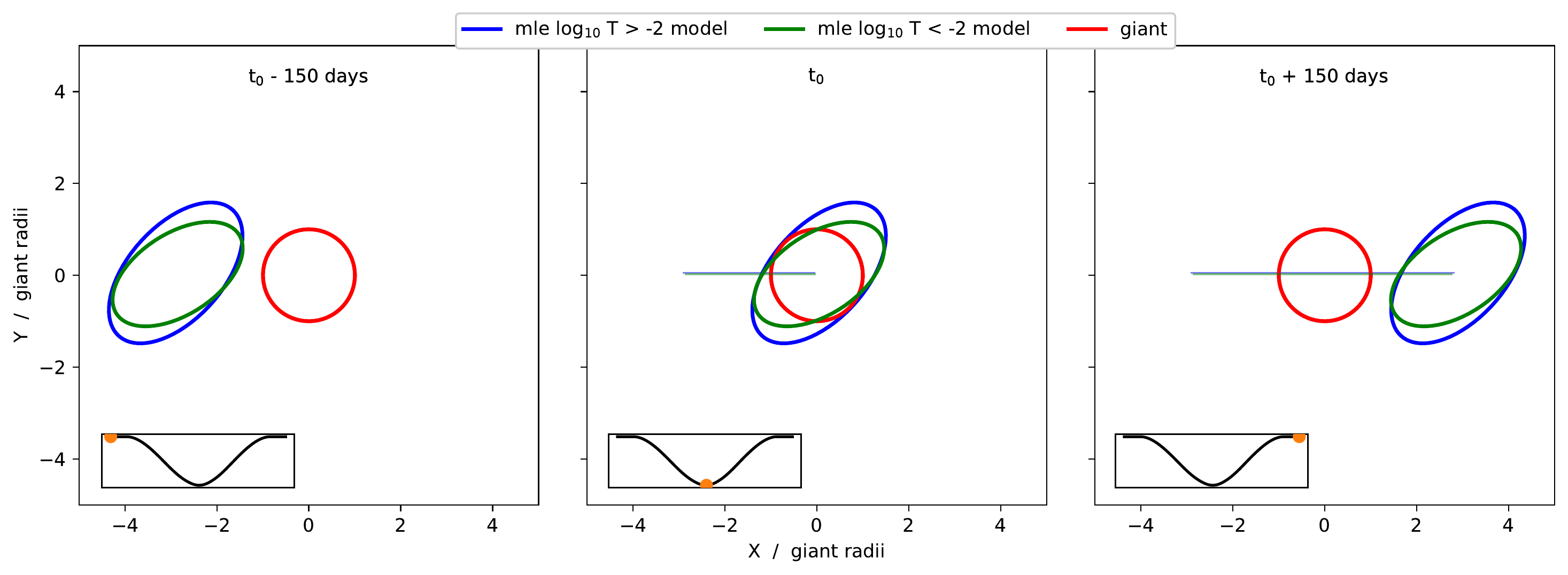}
    \caption{An illustration of the elliptical occulter eclipse models using maximum likelihood estimate (MLE) parameters for an opaque occulter ($\log_{10}~T<-2$, green) and a semi-transparent occulter ($\log_{10}~T>-2$, blue).}
    \label{occultation_cartoon}
  \end{center}
\end{figure*}

The posterior on the 9 parameters of the occultation model was sampled using nested sampling \citep{Skilling2004} via the pyMultinest package \citep{Feroz2009,pymultinest} with 5000 live points. The resulting posterior is shown on Figure~\ref{fig:occultation_corner}. We see that the inference is robust, with two distinct posterior modes corresponding to a completely opaque occulter, and an occulter with a $3$ per cent transparency. The opaque occulter requires that it is marginally smaller along one axis than the giant star, and implies that the remaining $3$ per cent flux in the light curve is due to a small un-obscured portion of the giant. The semi-transparent occulter requires that it is marginally larger along one axis than the giant, and implies that the remaining $3$ per cent flux in the light curve is due wholly to the transparency of the occulter. The two distinct transparency regimes result in two non-overlapping occulter orientation ($\theta$) modes. Generally speaking, the semi-transparent occulter allows more freedom across the remaining parameters.
A selection of model light curves sampled from the posterior is shown in Figure \ref{occultation_models}. The maximum likelihood models from each of the distinct modes are illustrated in Figure \ref{occultation_cartoon}.

{\renewcommand{\arraystretch}{1.5} 
\begin{table}
\begin{center}
\caption{Model parameters of the eclipse with priors and posterior measurements. Posterior column contains the median and 95th percentile credible interval in the parameters. Positions and proper motions are relative to those of the giant star, and are in units of $r_{\rm g}$.}
\label{model_params}
\resizebox{\columnwidth}{!}{
\begin{tabular}{|l|c|c|c|}
\hline
  \multicolumn{1}{|c|}{Parameter} &
  \multicolumn{1}{c|}{Prior} &
  \multicolumn{1}{c|}{Posterior} &
  \multicolumn{1}{c|}{Units} \\
\hline
$\log_{10}~r_{\rm maj}$    & ${\mathcal U}(-0.5, 2)$ $^a$     & $0.24^{+0.29}_{-0.05}$       & $r_{\rm g}$            \\
$\log_{10}~r_{\rm min}$    & ${\mathcal U}(-0.5, 2)$          & $0.004^{+0.010}_{-0.059}$    & $r_{\rm g}$            \\
$\theta{}$                 & ${\mathcal U}(0, \pi)$           & $0.79^{+0.22}_{-0.25}$       & radians                \\
$\log_{10}~r_{0}$          & ${\mathcal U}(-3, 2)$            & $-1.5^{+0.8}_{-0.3}$         & $r_{\rm g}$            \\
$\mu{}$                    & ${\mathcal U}(0.0, 0.2)$         & $0.0192^{+0.0011}_{-0.0007}$ & $r_{\rm g}$~day$^{-1}$ \\
$a$                        & ${\mathcal U}(0.5, 0.7)$         & $0.65^{+0.05}_{-0.13}$       &                        \\
$b$                        & ${\mathcal U}(0.05, 0.25)$       & $0.18^{+0.07}_{-0.12}$       &                        \\
$t_{0}$                    & ${\mathcal U}(54500, 57500)$     & $56021.2^{+5.0}_{-0.6}$      & MJD                    \\
$\log_{10}~T$              & ${\mathcal U}(-5, 0)$            & $-1.52^{+0.02}_{-3.39}$      &                        \\
\hline
\end{tabular}
}
\end{center}
$^a$ Also includes additional terms $r_{\rm maj}^{-2}$ and $r_{\rm min}/r_{\rm maj} \sim {\mathcal B}(1, 0.5)$, see text.
\end{table}
}

\subsection{Event chromaticity}\label{sec:chromaticity}

With a model of the event in hand we are now better able to quantitatively assess the chromatic nature of the event.

\begin{figure*}
  \begin{center}
    \includegraphics[width=\linewidth,keepaspectratio]{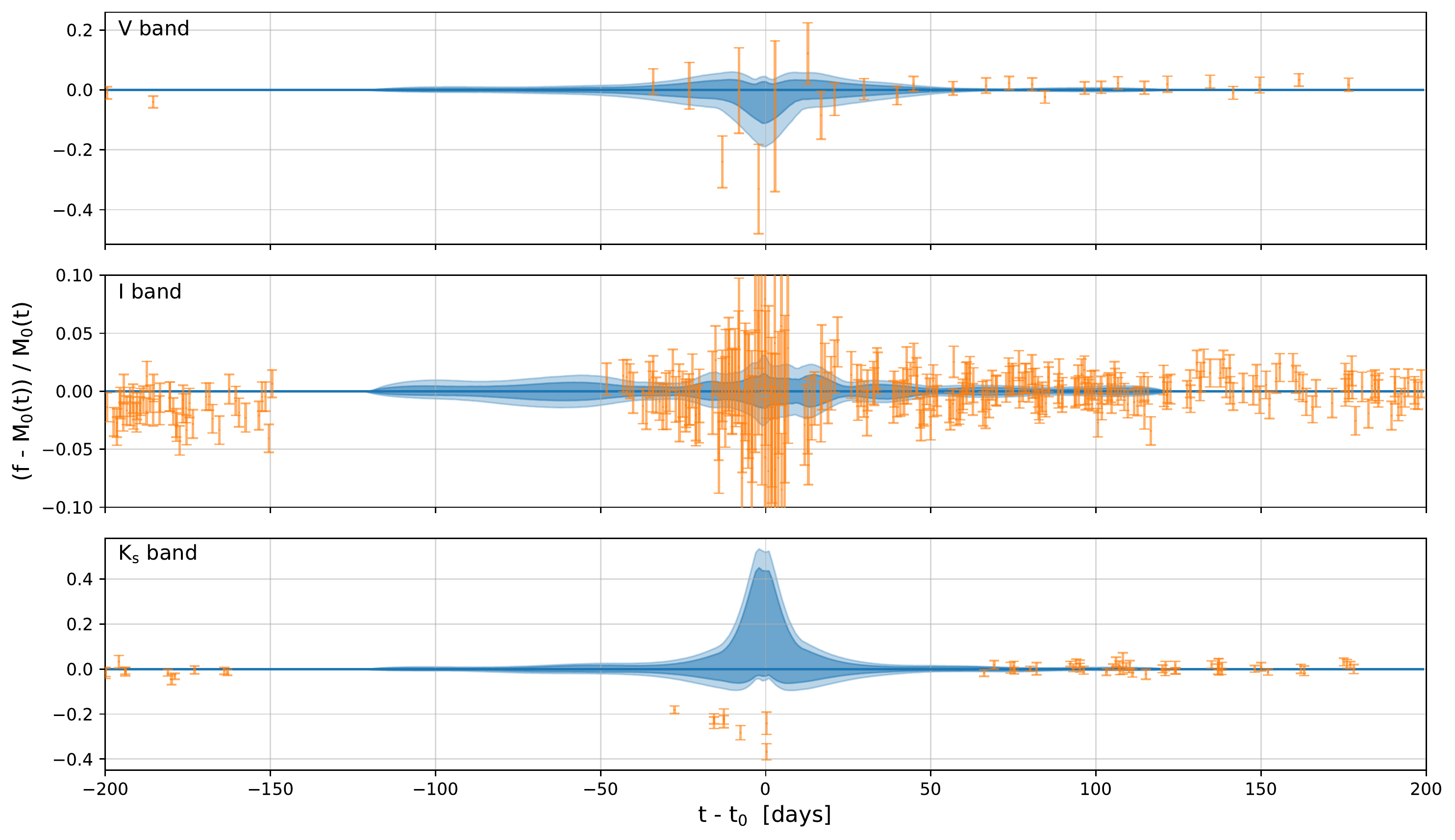}
    \caption{Fractional residuals to the median model flux for the $V$, $I$ and \Ks{} bands. The $V$ and \Ks{} models were obtained by substituting appropriate $V$ and \Ks{} band limb darkening parameters into the models obtained by fitting using the $I$ band data. The dark and light blue bands indicate 68th and 95th (respectively) percentile spread between the model samples.}
    \label{fig:chromaticity}
  \end{center}
\end{figure*}

Obtaining models derived from the $I$ band data but which are appropriate for the $V$ and \Ks{} bands simply requires substituting  the limb darkening parameters, while keeping the geometric parameters constant. If our assumptions that the event is achromatic and the $I$ band model are correct, then we should find that the $V$ and \Ks{} band data fit these modified models. By interpolating \citet{ld_bounds} limb darkening parameters we obtain $V$ and \Ks{} band equivalents to those of the $I$ band models, then we run these and the geometric parameters through the occultation modelling software described in Section \ref{modelling}. The resultant distributions of $V$ and \Ks{} band models are shown in Figure \ref{fig:chromaticity} against the observations.

The $V$ band data generally agrees with the $V$ band model prediction within the photometric errors and model spread, indicating no significantly different from the model $V-I$ colour. The same cannot be said for the \Ks{} band data, which is a few sigma below the prediction of the model between the start of the event and its peak. When the \Ks{} band observations resume towards the end of the event they are back where the model predicts, indicating a grey $I-K_s$ colour at this time. The $20$ to $30$ per cent lower \Ks{} band flux around the event peak evident in the lower panel of Figure \ref{fig:chromaticity} and is computed relative to the predicted model flux. If one instead considers the residual as a percentage of the baseline flux then it is $\sim{}1$ per cent.

We remark that the one further event candidate for which we have in-event data, \mbox{VVV-WIT-10}, also appears to exhibit a blue $I-K_s$ colour during the event. It is not detected in the OGLE $V$ band observations.

One possible explanation for the blue $I-K_s$ colour might be that flux contamination at baseline by a fainter, hotter object is masking the true obscuration fraction. A hot companion would contribute more flux in optical passbands, which might then not be obscured during the event, leading to an apparently shallower eclipse at shorter wavelengths. To test this hypothesis we reran the isochrone modelling described in Section \ref{isochrone}, this time included contamination from a fainter, hotter blackbody companion. The contaminating source is parametrised as a blackbody with a log-uniform $5000<T_{\rm eff}/{\rm K}<50000$ prior, and a contaminating flux fraction in the \Ks{} band (relative to the giant baseline flux) with the uniform prior between $0$ and $0.1$. The experiment revealed no clear evidence for such a contaminating component, although it is not ruled out with a high degree of confidence. It is unfortunate we do not have sufficient data to conduct this test with the in-event photometry. With the baseline photometry we are trying to identify a signal which is approximately the same size as the typical photometric uncertainties.

Weak evidence in favour of the flux contamination scenario is the presence of the $3$ per cent transparency mode in the event modelling posterior, which is directly interchangeable with flux contamination. That is, the measured $3$ per cent transparency might instead be $3$ per cent flux contamination in the $I$ band. Further evidence for the flux contamination scenario is that the colour change is expected to be proportional to the fraction of obscured giant flux. It therefore is able to explain the return to a zero (or near-zero) $I-K_s$ colour at egress. Note that the contaminant need not be the occulter itself, it could be another member of the system or an unrelated object along the line of sight.

A second, discounted, hypothesis was that an opacity hole in the center of the occulting disc would allow light in from the central region of the disc of the background giant. In our occultation model the center of the occulting disc has a relatively low impact parameter in some cases. As flux is concentrated more towards the limb of the background giant in the near-infrared this would manifest as a deeper eclipse at these wavelengths. The counterargument to this theory is that the opposite effect is expected to be seen at shorter wavelengths, i.e. where the \Ks{} eclipse would be deeper than the $I$ band, the $V$ ought to be shallower, yet this is not the case. Furthermore, the discontinuity in $I-K_s$ colour between the early and late eclipse cannot be explained through this method.

Another possible explanation for the blue $I-Ks$ colour may be that some scattered light from the edges of the occulter was detected at mid-eclipse. The single scattering albedo of sub-micrometre dust grains is higher in the optical waveband than in the infrared, see e.g. \citet{Lucas98}. The phase function of such grains also causes an increase in forward scattering in the optical, relative to the infrared. Therefore, if the achromatic nature (to first order) of the event is due mainly to high optical depth rather than a large grain size distribution, then forward scattering of light from the giant star by the edges of an occulting disc can plausibly explain the slightly greater dimming in the infrared. This explanation is invalid if the disc is transparent to some degree, and while the modelling tends to prefer a $3$ per cent transparency, this is not a given (as discussed in Section \ref{sec:discuss_event_modelling}).

\section{Discussion}\label{discussion}

\subsection{Event modelling}\label{sec:discuss_event_modelling}

\begin{figure}
  \begin{center}
    \includegraphics[width=\linewidth,keepaspectratio]{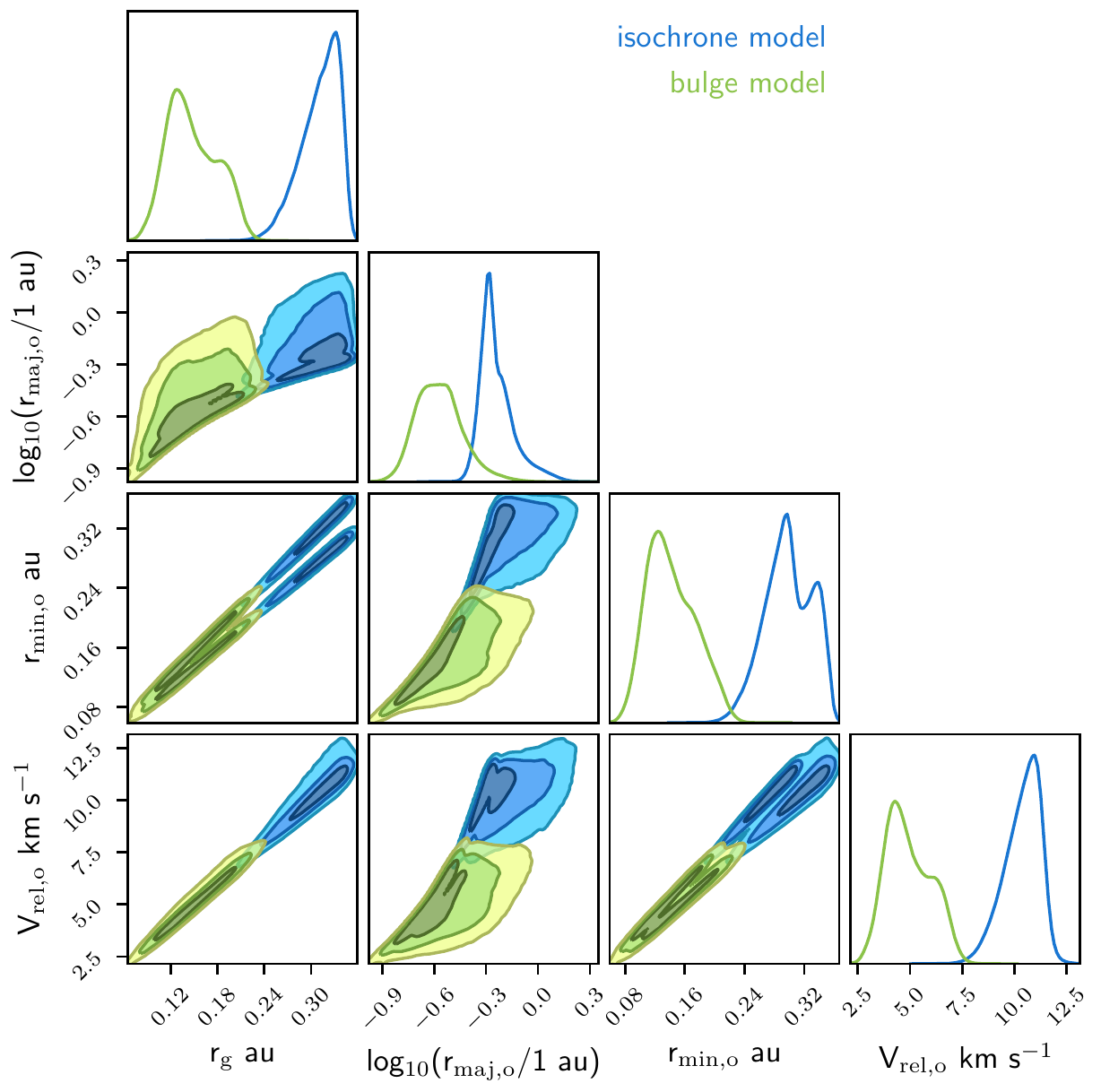}
    \caption{Posterior pdf of giant radius and occulter size and velocity for the two giant models considered: one obtained from isochrone fitting (isochrone model), and one obtained assuming a bulge location for the giant (bulge model).}
    \label{occulter_params}
  \end{center}
\end{figure}

Figure \ref{occulter_params} shows the posterior pdf on occulter parameters implied by the event modelling for the two giant star models considered: the isochrone-based parameters with the $V_{\rm tan}(R)<V_{\rm esc}(R)$ requirement (Section \ref{isochrone}, `isochrone model' hereafter), and the parameters implied by requiring the giant is located in the bulge (Section \ref{BB_Bulge_rad}, `bulge model' hereafter). Further details are provided in Table \ref{tab:occulter_params}.

The relative velocity obtained from the modelling of our occultation is well constrained by the data and strongly covariant with the radius of the giant star. If we assume a circular orbit of the occulter around the giant then the only remaining free parameter in the determination of orbital separation and period is the combined mass of the system. The orbital period and radius per $M_{\odot}$ of system mass for the \wit{} binary system are given in Table \ref{tab:occulter_params}. We obtained an estimate of the mass of the giant in Section \ref{isochrone} of $\sim{}1~M_{\odot}$ for the isochrone model. If the giant has a smaller radius than measured from the isochrones, as in the bulge model, then this requires lower mass to maintain the value of $\log{g}$ measured from the NIR spectrum. The mass implied for the giant star in the bulge model is of order $0.4~M_{\odot}$, with large error bars due in part to the large assumed systematic error on $\log{g}$.

As noted in Section \ref{modelling}, the inferred upper limit on major axis radius and axis ratio are dictated largely by the prior. Only the minor axis radius is strongly constrained by the data, since the depth is almost 100 per cent and the light curve is not flat-bottomed the radius of the occulter on its shortest dimension must be approximately equal to the radius of the giant. This caveat should be kept in mind when considering the size of the occulter. The lower bound of the inferred major axis radius is constrained to a reasonably high degree of confidence, due to the need for it to be somewhat larger than the minor axis radius, since occulters that are circular in projection cannot produce the observed light curve (as seen in Figure \ref{occultation_models}).
Relatedly, it's possible that we are instead seeing an edge-on disc and that what we had thought of as the minor axis radius is instead its thickness. A conservative estimate of the maximum size of the disc is the orbital separation of the system, which is system mass and giant radius dependent.

{\renewcommand{\arraystretch}{1.5} 
\begin{table}
\begin{center}
\caption{Predicted occulter parameters for the two giant star scenarios we considered. `bulge model' is a sub-luminous giant star at the distance of the Galactic bulge, and `isochrone model' is a giant star at $\approx{20}$~kpc with an improbably large space velocity. We give the median and $95$ per cent confidence interval for each parameter. Subscripts `g' and `o' refer to the giant star and occulter respectively. Most of the parameters are self explanatory. Min M$_{K_s\rm{},o}$ is an approximate minimum $K_s$ band absolute magnitude of the occulter under the assumptions (i) that it completely obscures the giant and therefore constitutes $3$ per cent of the combined flux of the giant and occulter, and (ii) that the occulter is not itself obscured. The orbital period and separation axis estimates P$_{\rm orbit}$ and a$_{\rm orbit}$ are per $M_{\odot}$ of system mass, under the assumption of a circular orbit.}
\label{tab:occulter_params}
\begin{tabular}{|l|c|c|c|}
\hline
  \multicolumn{1}{|c|}{Parameter} &
  \multicolumn{1}{c|}{bulge model} &
  \multicolumn{1}{c|}{isochrone model} &
  \multicolumn{1}{c|}{Units} \\
\hline
r$_{\rm{}g}$                 &  $0.14^{+0.07}_{-0.05}$   &  $0.32^{+0.03}_{-0.06}$   &  au     \\
r$_{\rm{}maj,o}$             &  $0.27^{+0.32}_{-0.11}$   &  $0.54^{+0.64}_{-0.12}$   &  au     \\
r$_{\rm{}min,o}$             &  $0.14^{+0.07}_{-0.05}$   &  $0.30^{+0.05}_{-0.06}$   &  au     \\
V$_{\rm{}rel,o}$             &  $4.8^{+2.3}_{-1.6}$      &  $10.5^{+1.2}_{-2.0}$     &  km~s$^{-1}$  \\
P$_{\rm{}orbit}$/$M_{o+g}$   &  $242^{+597}_{-166}$      &  $23^{+21}_{-6}$          &  years~$M_{\odot}^{-1}$  \\
a$_{\rm{}orbit}$/$M_{o+g}$   &  $39^{+50}_{-21}$         &  $8^{+4}_{-2}$            &  au~$M_{\odot}^{-1}$  \\
min M$_{K_s\rm{},o}$         &  $0.34^{+0.89}_{-0.83}$   &  $-1.32^{+0.44}_{-0.18}$  &  mag  \\
\hline
\end{tabular}
\end{center}
\end{table}
}

The $I$ band event depth as measured from the models is $96.93^{+0.05}_{-0.06}$ per cent at $1\sigma$, the maximum possible transparency of the occulter is therefore $3$ per cent, which is the upper limit dictated by the event modelling. This is inconvenient as achromaticity is easier to explain with a high optical depth, but we note that swapping occulter transparency for flux contamination will yield the same light curve. As noted previously, flux contamination may have evaded our test in Section \ref{sec:chromaticity} due to the signal being of roughly the same magnitude as the photometric uncertainties. Additionally, the contaminating flux test did not include contaminants with $T_{\rm eff}<5000$~K, leaving room for a cooler companion. Furthermore, there is plenty of room in the event model posterior for opaque occulters.

\subsection{The nature of the occulter}

We have demonstrated that a solid occultation model fits the available data. We cannot rule out an occulter with a radial transparency gradient, for example, but this could not explain the relative achromaticity across wavelengths of $0.5$ to $2.2$~$\mu{}m$ unless it were comprised principally of large-grained material. Further investigation, including modelling of the event with such an occulter would still be worthwhile.

As noted in Sections \ref{isochrone} and \ref{BB_Bulge_rad}, the measured radius and/or space velocity of \wit{} are problematic to some degree. If the giant is on an ordinary single-star evolutionary pathway then it has an unusually large space velocity, whereas if it has a typical space velocity and distance for objects in this direction then it has an unusually small radius. It is possible there is a middle ground to be found between the two scenarios. An explanation for the event which also offers an explanation for the apparent unusual nature of \wit{} is desirable.

The required features for the occulter are that
\begin{inparaenum}
\item it is bound to the giant,
\item it is faint,
\item it has a radius greater than $0.25$~au,
\item it is elliptical in projection.
Additional desirable features are that
\item its presence also explains a high space velocity or smaller giant radius,
\item it can explain a mildly blue $I-K_s$ colour during the first half of the event.
\end{inparaenum}

Based on these requirements, in particular requirement (iv), it is likely the occulter is a circumstellar disc. A circumstellar disc could be hosted by a body which is gravitationally bound to the giant, and its hosting body could be faint or itself obscured by the disc. Dusty circumstellar discs are observed around all phases of stellar evolution~\citep{Zuckerman2001}. We discuss them in turn.

\subsubsection{Pre-main-sequence disc}

We consider the possibility of occultation by a pre-main-sequence or YSO disc. Brown dwarf discs found in star forming regions typically have radii ranging from a few au to several tens of au \citep[e.g.][]{2006ApJ...645.1498S,2017ApJ...841..116H,2019ApJ...878..103R}, which is consistent with the inferred radius of the occulting disc in our model.

One notable example is Mamajek's object, a $\sim{}16$~Myr old pre-main sequence star that exhibited a long, deep and complex dipping event which is consistent with occultation by a series of rings akin to the densest B-ring of Saturn \citep{Mamajek12}. These rings are hosted by an unobserved companion body, likely to be a brown dwarf or a massive planet.

For Mamajek's object the outermost ring is approximately $0.6$ au in radius and, assuming similar properties to Saturn's B-ring, the entire system must contain $\sim100~M_\mathrm{Moon}$ of material \citep{Kenworthy2016}. The opacity of the rings appears to increase inwards and it is believed the sequence of rings was potentially carved out by exomoons. It is plausible that our occulter is similar to the disc in Mamajek's object, based on the size and the range of opacities, and so would be associated with a similar mass of material. Since for our object the light curve is smooth the occulter cannot be as structured as Mamajek's.

YSO accretion discs have a typical lifetime of a only few Myr. We would not expect to find a bound YSO companion to the giant unless the giant were very massive, distant and with ${\rm V}_{\rm tan}\gg{\rm V}_{\rm esc}$. Furthermore, if we adopt a bulge location for the giant star, then the occulter is at a heliocentric distance of at least $6$~kpc and therefore greater than $400$~pc below the plane of the Milky Way. The distance from the plane is naturally larger for the isochronal giant model with a larger heliocentric distance. The scale height of molecular clouds and pre-main sequence stars in the Milky Way is only $35$ to $50$~pc \citep[e.g.][]{2005ApJ...619L.159S, 2016BaltA..25..261B}, so such a youthful object is very unlikely to be found in such a location.

The difficulties with the pre-main sequence disc scenario continue with the colour of the event. Extinction by optically thick YSO discs is generally not achromatic: at optical and near infrared wavelengths sub-micrometre-sized grains are generally observed \citep{2006ApJ...638..314D}. 
Larger grains settle to the disc's mid-plane and are observed only at far-infrared and submillimetre wavelengths. Grey extinction by larger grains has occasionally been suggested at optical and near infrared wavelengths \citep[e.g.][]{2001Sci...292.1686T,2016MNRAS.463.4459B} but to our knowledge it has never been confirmed by follow-up work \citep[see e.g][]{Miotello12}.

On consideration of the above we conclude that our occulter cannot be a disc hosted by a pre-main-sequence star.

\subsubsection{Main-sequence discs}

Large and optically thick discs hosted by main sequence stars are generally confined to those that are relatively young, and these can be discounted using the same scale height argument as for pre-main-sequence stars.

Older main-sequence stars can host debris discs, these were first observed around Vega and are analogous to the Solar System's Kuiper belt. Infrared excesses suggest debris discs are present around many main sequence stars \citep{Hughes2018}. These discs extend up to $10-200$~au from the host star, and since the upper limit on occulter $r_{\rm maj}$ is poorly constrained by our data we cannot rule out this scenario on these grounds alone. However, main-sequence star debris discs are optically thin, so we consider an occultation by a debris disc hosted by a companion main-sequence star to be an unlikely scenario.

\subsubsection{White dwarf debris disc}

\cite{Farihi2016} reviews the evidence for circumstellar discs around white dwarfs. The pollution of metals in white dwarf atmospheres suggest many of these objects experience accretion. \cite{Jura03} hypothesised the accretion could be from an optically-thick disc of dust formed from the tidal disruption and subsequent grinding down of asteroids. Such a picture requires the disc to extend no further than the white dwarf's Roche radius of $\sim 1 R_\odot$, which is corroborated by infra-red excess measurements \citep[e.g.][]{Jura2007}. Furthermore, several white dwarfs have been observed to have double-peaked CaII emission lines indicative of a gaseous component in the surrounding disc constraining the outer radius to $\sim 1~R_\odot$ \cite[e.g.][]{Gaensicke2006}. 

Whilst typical white dwarf debris discs disrupted at the Roche limit are too small, the expansion of a red giant causes major disruption of the associated solar system due to the loss of mass from the system. This might lead to increased collisions amongst the remaining asteroids and planets in a solar system \citep[see modelling by][]{2020MNRAS.497.4091M} that could perhaps cause the formation of a much larger white dwarf debris disc than has been observed hitherto, but this is mere speculation.

From consideration of the timescale of Poynting-Robertson drag on small particles and the observed atmospheric pollution, it is understood that the discs must be optically thick \citep{Rafikov2011}, although \cite{Bonsor2017} argue this may be inconsistent with the frequency of infrared excesses observed in an unbiased sample of white dwarfs. Finally, theoretical models suggest that sublimation occurs around $0.2R_\odot$ \citep{Rafikov2011} producing an inner hole, whilst the outer edge of the disc is sharply truncated as optically thin material is rapidly dragged into the higher density regions \citep{BochkarevRafikov}. The event modelling suggests that the impact parameter $r_0$ is always smaller than $r_{\rm g}$, i.e. the center of the occulter passes over the stellar disc. However, the log-uniform prior on $r_0$ tends to restrain it to smaller values. It is possible there is room for models with large $r_{\rm maj}$ and $r_0$. Such models could have an inner hole in the occulting disc which would not impact the observed light curve.

Based upon these considerations, although the optical thickness, hard edge and low luminosity favour a white dwarf debris disc, they do not match the observed size. The conservative minimum size of the occulter is $50~R_{\odot}$. We therefore conclude a debris disc hosted by a white dwarf is an unlikely potential occulter.

\subsubsection{Post-main-sequence -- B-type subdwarf disc}

B-type subdwarf (sdB) stars are core helium burning stars at the blue end of the horizontal branch. Their existence can be explained through binary population synthesis models, around half are found in close binaries with white dwarfs or low-mass main-sequence stars \citep{heber09}. 

The scenario favoured by \citet{Rodriguez16} in the case of TYC 2505-672-1 -- the most similar example to \wit{} that we have found in the literature -- is of an sdB star hosting a large and opaque disc. In their case, UV observations by GALEX \citep{bianchi11} facilitated the identification of the sdB star in the composite SED. The GALEX All-sky Imaging Survey and Medium-depth Imaging Survey cover \wit{} but neither resulted in a detection at its position. Short wavelength observations are hampered by the relatively high extinction in this direction in any case.

While we consider this a plausible scenario for \wit{}, our concern is that the formation of an sdB star typically requires mass transfer from the sdB progenitor to a companion. If this were the case, it would make an apparent reduction in the size of the giant star even harder to explain, though perhaps a suitable explanation can be found should further observation reveal the presence of such a companion.

\subsubsection{Disc comprised of matter stripped from the giant}\label{sec:strippeddisc}

We speculate that mass transfer from the giant to the gravitational influence of a companion with sufficient angular momentum might explain the presence of a disc. This scenario is attractive as it offers a convenient explanation of the potentially smaller than expected radius of the giant. Any plausible companion of the giant star might be a host to such a disc -- a main sequence star, itself a close binary system, a white dwarf or other compact object -- but there are some additional considerations.

If the companion is on the main-sequence it naturally must have been less massive than the giant during its main-sequence in order for the giant to have evolved faster. Most main-sequence stars less massive than the giant are fainter than our detection threshold even if they are not themselves obscured. We note that the canonical $\epsilon{}$ Aurigae scenario is of a born-again post-AGB F-type supergiant primary that has a B-type main sequence companion hosting a large, opaque disc. While the spectral types of the stars are different and the disc is bright at long wavelengths \citep{Hoard10}, the $\epsilon{}$ Aurigae system resembles \wit{} in many ways. An explanation for the lack of a mid-infrared excess in the AllWISE and GLIMPSE photometry (see Figure \ref{sed_model}) of \wit{} might simply be that the disc is too small. The surface brightness of $3600$~K blackbody (the giant star) is $\sim{}50\times{}$ larger than that of a $500$~K blackbody in the Spitzer $8.0~\mu{m}$ band, and $\sim{}150\times{}$ brighter in the $5.8~\mu{m}$ band. With the GLIMPSE uncertainties of $\sim{}0.05$~mag for \wit{} we would not expect to be able detect the presence of a $500$~K disc at $8.0~\mu{m}$ if it has an angular area smaller than $\sim{}2.5\times{}$ that of the giant. This not considering the potential for contamination in the relatively low resolution mid-infrared data due to the close proximity of unrelated neighbouring stars on the sky.

Another consideration is that our estimate of the \textit{current} orbital separation of the two components is too large to facilitate Roche lobe overflow. Our predicted orbital separation per $M_{\odot}$ of system mass is given in Table \ref{tab:occulter_params}. Using the posterior on the mass and radius of the giant from Sections \ref{isochrone} and \ref{BB_Bulge_rad}, and the \citet{eggleton83} approximation to the Roche lobe radius we find the closest the giant comes to filling its Roche lobe is $\sim{}10$~per~cent with $M_{\rm giant}/M_{\rm occulter}=q\approx{}1$ for companion masses between $10^{-2}$~$M_{\odot}$ and $10$~$M_{\odot}$. The more massive the companion, the smaller the filled fraction of the Roche lobe radius of giant star, due to the larger orbital separation required to explain the velocity of the occulter relative to the giant. Hence, unless the companion is on a highly elliptical orbit, outward migration due to mass loss from the system is probably required.

We should also consider that the event modelling prefers an occulter with a position angle that is significantly offset from the direction of velocity ($\theta{}\approx{}60^{\circ}$). If the disc were comprised of material stripped from the giant we might expect it to be aligned with the orbital plane of the companion. This though might be explained through a companion with an inclined rotation axis relative to its orbital plane and a magnetic field facilitating transfer of angular momentum to the accreted material, for example.

\subsubsection{Black hole or neutron star disc}

Related to Section \ref{sec:strippeddisc}, one possibility might be an accretion disc around a black hole or neutron star companion. \citet{La16} estimates the size of black hole accretion discs as up to $10^6$ Schwarzschild radii. A disc of minimum size $0.25$~au therefore requires a black hole with a minimum mass of approximately $\sim{}10~M_{\odot}$. If we adopt the more realistic $1$~au minimum occulter radius then the minimum mass increases to $\sim{}50~M_{\odot}$. Such accretion discs are likely optically thick but will usually emit X-rays, yet no sources are present within 2\arcmin of \wit{} in either the XMM-Newton Serendipitous Source Catalogue \citep[3XMM-DR8;][]{3xmm} or the Second ROSAT all-sky survey (2RXS) source catalog \citep{2rxs}. Even if quiescent now, quiet periods are punctuated by outbursts, so the object would nonetheless be a transient X-ray source. However, there are no very sensitive X-ray observations of this field, so it is conceivable that such an object might have been missed.

We discuss in Section \ref{sec:strippeddisc} the increase in the predicted orbital separation of the \wit{} system components that is caused by such a massive companion, and hence the inability of the giant to fill its Roche lobe. This then requires a quiescent accretion disc that is formed of material stripped from the giant prior to a radial migration, or is formed of remnant material in the vicinity of the companion due to mass loss during its evolution to a compact object. The latter scenario leads us to considering discs formed from black hole fallback material of the type described by \citet{perna14} and elaborated on in the context of stellar occultation by \citet{Wright16}. Such a disc is a plausible occulter in the \wit{} system. We note that black hole fallback discs require an inner hole, and this is not necessarily ruled out by our occulter modelling.

\subsection{Contextualising VVV-WIT-08}

In addition to \wit{} we have identified two further promising candidate events in our VVV and VVVX data based on shallow searches, \mbox{VVV-WIT-10} and \mbox{VVV-WIT-11}. These additional candidates reside on the giant branch of the CMD. \wit{} is a confirmed late-type giant star, the only open question being its distance and hence radius. The analyses of TYC~2505-672-1 by \citet{Rodriguez16}, and ASASSN-21co by \citet{asassn-21co} indicate that they too are late-type giant stars. While the $\epsilon{}$ Aurigae primary is a somewhat hotter F0 star with supergiant surface gravity.
We speculate that long-term, wide-field stellar photometric monitoring is beginning to reveal the presence of a population of long-period eclipsing binaries composed of giant stars and opaque-disc-hosting companions.

\section{Conclusions}\label{conclusion}

We have reported and analysed an unusual dipping event in the light curve of a late-type giant star, \wit{}. The dip is unusually long at over a few hundred days and has a depth of $97$ per cent. It is largely achromatic, though we observe a mildly bluer than expected $I-K_s$ colour during the first half of the event.

The object was found serendipitously in the VISTA Variables in the Via-Lactea (VVV) survey. Subsequent searches revealed two further event candidates, \mbox{VVV-WIT-10} and \mbox{VVV-WIT-11}, although their light curves are much less well sampled. \citet{cp17a} and Lucas et al. (in preparation) find a number of long duration `dipper' events associated with Young Stellar Objects in VVV searches for high amplitude variable sources, but none share the distinctive smooth dimming of \wit{}.

The proper motion of \wit{} is entirely consistent with a location in the Galactic bulge. However, the fitting of isochrones to the photometry gives a larger distance of $\approx{}20$ kpc. Such a distance seems relatively unlikely, as the inferred space velocity is then comparable or larger than the Galactic escape speed. It is possible that \wit{} is relatively underluminous, perhaps a sub-subgiant star or other product of binary stellar evolution, which renders the isochronal fit suspect.

The behaviour of the light curve of \wit{} does not correspond to any known intrinsic stellar variability. It is almost certainly an occultation of the giant star. We find that an occultation due to a chance alignment with the giant star requires an improbably large space density of dark foreground objects. Hence the occulter is most likely gravitationally bound to the giant. By modelling the light curve as an occultation by an object that is elliptical in projection and has a uniform transparency, we find that the occulter must be faint, likely optically thick and possess a radius or thickness in excess of $0.25$ au. The inferred orbital parameters of the system weakly constrain the maximum size of the occulter to tens or hundreds of au, dependent on the true radius of the giant and the mass of the system. Our modelling implies that a hard edge of the occulter is necessary. We justify this on the grounds that it is more difficult to explain the relative achromaticity with an opacity gradient, but modelling of such a system is encouraged nevertheless.

We speculate that \wit{}, the two additional candidates, \mbox{TYC~2505-672-1} and \mbox{ASASSN-21co} might be representative of a population of long-period eclipsing binaries composed of late-type giant stars and opaque-disc-hosting companions. $\epsilon{}$ Aurigae might be considered an additional member if the spectral types of its components can be overlooked. If this population exists, future wide-field time series photometric surveys such as the {\it Legacy Survey of Space and Time} (LSST) at the Vera Rubin Observatory~\citep{Sa19} may reveal more such objects.

We considered a number of possible astrophysical objects as candidates for the occulter. Debris discs around main sequence stars are too optically thin. Whilst white dwarf debris discs are optically thick, they are too small. Accretion discs around black holes and neutron stars usually emit X-rays, but a black hole fallback disc of the type described by \citet{perna14} might plausibly explain the occultation. A disc formed of material stripped from the giant is an attractive scenario. A difficulty with this is that the present orbital separation cannot facilitate Roche-lobe overflow. However, this could be explained by an increase in orbital radius due to mass loss from the system. An optically thick YSO disc or an analogue of Mamajek's object are approximately the right size and could be optically thick. However, the height of the object above the Milky Way disc plane argues against both possibilities, given the scale height of young populations. Despite intensive efforts, it is clear that we have left room for further work on this intriguing object!

\section*{Acknowledgements}

Thanks are due to the anonymous referee whose comments helped strengthen this analysis significantly, and to J.E. Drew for supplying VPHAS+ data and suggestions regarding the nature of \wit{}.

The OGLE project has received funding from the National Science Centre, Poland, grant MAESTRO 2014/14/A/ST9/00121 to AU.

This work has made use of the University of Hertfordshire's high-performance computing facility.

J.L.S. acknowledges support from the Royal Society (URF\textbackslash R1\textbackslash191555). 
D.M. gratefully acknowledges support by the BASAL Center for Astrophysics and Associated Technologies (CATA) through grant AFB 170002, and by Proyecto FONDECYT regular No. 1170121.
R.K.S. acknowledges support from CNPq/Brazil through project 305902/2019-9.

The authors wish to acknowledge their use of following data sources, services and software packages:

The Southern Astrophysical Research (SOAR) telescope, which is a joint project of the Minist\'{e}rio da Ci\^{e}ncia, Tecnologia, Inova\c{c}\~{o}es  (MCTI) do Brasil, the US National Science Foundation's NSF's NOIRLab (NOIRLab), the University of North Carolina at Chapel Hill (UNC), and Michigan State University (MSU).

The European Space Agency (ESA) mission
{\it Gaia} (\url{https://www.cosmos.esa.int/gaia}), processed by the {\it Gaia}
Data Processing and Analysis Consortium (DPAC,
\url{https://www.cosmos.esa.int/web/gaia/dpac/consortium}). Funding for the DPAC
has been provided by national institutions, in particular the institutions
participating in the {\it Gaia} Multilateral Agreement.

The VizieR catalogue access tool, CDS, Strasbourg, France (DOI: 10.26093/cds/vizier). The original description of the VizieR service was published in \citet{2000A&AS..143...23O}.

The SVO Filter Profile Service (\url{http://svo2.cab.inta-csic.es/theory/fps/}) supported from the Spanish MINECO through grant AYA2017-84089

The Whole Sky Database (wsdb) created by Sergey Koposov and maintained at the Institute of Astronomy, Cambridge by Sergey Koposov, Vasily Belokurov and Wyn Evans with financial support from the Science \& Technology Facilities Council (STFC) and the European Research Council (ERC).

Astropy,(\url{http://www.astropy.org}) a community-developed core Python package for Astronomy \citep{astropy:2013, astropy:2018}.

The ChainConsumer \citep[\url{https://samreay.github.io/ChainConsumer/};][]{Hinton2016}, IMF (\url{https://github.com/keflavich/imf}), and Schwimmbad \citep[\url{https://github.com/adrn/schwimmbad};][]{schwimmbad} python packages, and the AGAMA library for galaxy modelling \citep[\url{https://github.com/GalacticDynamics-Oxford/Agama};][]{agama}.

\section*{Data Availability}
Light curves for \mbox{VVV-WIT-08}, \mbox{VVV-WIT-10}, and \mbox{VVV-WIT-11} are published with this article as supplementary material.

The light curve model fitting code is available at \url{https://github.com/segasai/vvv_dipper_code}.



\bibliographystyle{mnras}
\bibliography{refs}





\bsp	
\\

\label{lastpage}
\end{document}